\documentstyle[]{mn}
\input{psfig}
\def\modF0{{\tt Fiducial}}
\def\modG0{{\tt Radial}}
\def\modD10{{\tt Poisson}}
\def\modH0{{H0}}
\def\modI0{{I0}}
\def\modJ0b{{\tt noRSSD}}
\def\modM0{{\tt MassBias}}
\def\modK0{{\tt noSGW}}

\def\beq{\begin{equation}}
\def\eeq{\end{equation}}
\def\barray{\begin{eqnarray}}
\def\earray{\end{eqnarray}}
\def\beqarray{\begin{eqnarray}}
\def\eeqarray{\end{eqnarray}}

\def\rmb{{\rm b}}
\def\rmc{{\rm c}}
\def\rmd{{\rm d}}

\def\rmg{{\rm g}}
\def\rmh{{\rm h}}

\def\rmm{{\rm m}}

\def\rmp{{\rm p}}

\def\rms{{\rm s}}

\def\rmx{{\rm x}}
\def\rmy{{\rm y}}

\def\rmP{{\rm P}}

\def\calA{{\cal A}}

\def\calD{{\cal D}}

\def\calG{{\cal G}}
\def\calH{{\cal H}}

\def\calL{{\cal L}}

\def\calP{{\cal P}}

\def\calR{{\cal R}}

\def\bw{{\bf w}}
\def\bx{{\bf x}}

\def\bA{{\bf A}}

\def\bC{{\bf C}}

\newcommand{\etal}{{et al.~}}

\newcommand{\kmsmpc}{\>{\rm km}\,{\rm s}^{-1}\,{\rm Mpc}^{-1}}

\newcommand{\Mpc}{\>{\rm Mpc}}

\newcommand{\Msun}{\>{\rm M_{\odot}}}

\newcommand{\Ombh}{\Omega_{\rm b}\,h^{2}}
\newcommand{\bolds}[1]{{\bf #1}}

\def\gtsima{$\; \buildrel > \over \sim \;$}
\def\ltsima{$\; \buildrel < \over \sim \;$}
\def\prosima{$\; \buildrel \propto \over \sim \;$}
\def\gsim{\lower.7ex\hbox{\gtsima}}
\def\lsim{\lower.7ex\hbox{\ltsima}}
\def\simgt{\lower.7ex\hbox{\gtsima}}
\def\simlt{\lower.7ex\hbox{\ltsima}}
\def\simpr{\lower.7ex\hbox{\prosima}}

\def\ga{\gsim}

\def\gta{\ga}

\newcommand{\apj}{ApJ}
\newcommand{\apjs}{ApJS}

\newdimen\hssize
\newdimen\hdsize
\begin{document}
%

\title[Cosmological Constraints from Clustering and Lensing]
      {Cosmological Constraints from a Combination of Galaxy Clustering 
         and Lensing -- III. Application to SDSS Data}

\author[Cacciato et al.]
       {\parbox[t]{\textwidth}{
        Marcello Cacciato$^{1}$\thanks{
           E-mail: cacciato@phys.huji.ac.il}, 
        Frank C. van den Bosch$^2$, 
        Surhud More$^{3}$,
        Houjun Mo$^4$,\\
        Xiaohu Yang$^5$} \\ 
           \vspace*{3pt} \\
	$^1$Racah Institute of Physics, The Hebrew University,
            Jerusalem 91904, Israel\\
	$^2$Department of Astronomy, Yale University, 
            P.O. Box 208101, New Haven, CT 06520-8101\\
	$^3$Kavli Institute for Cosmological Physics, University of Chicago, 
	     933 East 56th Street, Chicago, IL 60637, USA\\
        $^4$Department of Astronomy, University of Massachusetts,
            Amherst MA 01003-9305\\
        $^5$Key Laboratory for Research in Galaxies and Cosmology, Shanghai
            Astronomical Observatory, the Partner Group of MPA,\\
        $\;$Nandan Road 80, Shanghai 200030, China}


\date{}

\pagerange{\pageref{firstpage}--\pageref{lastpage}}
\pubyear{2008}

\maketitle

\label{firstpage}


\begin{abstract}
  We simultaneously constrain cosmology and galaxy bias
  using measurements of galaxy abundances, galaxy clustering and
  galaxy-galaxy lensing taken from the Sloan Digital Sky Survey.
  We use the conditional luminosity function (which
  describes the halo occupation statistics as function of galaxy
  luminosity) combined with the halo model (which describes the
  non-linear matter field in terms of its halo building blocks) to describe 
  the galaxy-dark matter connection. We explicitly account for  residual redshift
  space distortions in the projected galaxy-galaxy correlation functions,
  and marginalize over uncertainties in the scale
  dependence of the halo bias and the detailed structure of dark
  matter haloes. Under the assumption of a spatially flat, vanilla
  $\Lambda$CDM cosmology, we focus on constraining the matter density, 
  $\Omega_\rmm$, and the normalization of the
  matter power spectrum, $\sigma_8$, and we adopt WMAP7 priors 
  for the spectral index, $n_\rms$, the Hubble parameter, $h$, 
  and the baryon density, $\Omega_{\rm b}$. We obtain that $\Omega_\rmm = 0.278_{-0.026}^{+0.023}$ and
  $\sigma_8 = 0.763_{-0.049}^{+0.064}$ (95\% CL). These results are
  robust to uncertainties in the radial number density distribution of
  satellite galaxies, while allowing for non-Poisson satellite
  occupation distributions results in a slightly lower value for
  $\sigma_8$ ($0.744_{-0.047}^{+0.056}$). These constraints
  are in excellent agreement (at the $1\sigma$ level) with the cosmic 
  microwave background constraints from WMAP.
  This demonstrates that the use
  of a realistic and accurate model for galaxy bias,
  down to the smallest non-linear scales currently observed in galaxy surveys,
  leads to results perfectly consistent with the vanilla $\Lambda$CDM cosmology.
\end{abstract}




\section{Introduction}
\label{sec:intro}

The last two decades have seen the emergence of a concordance
cosmological model which describes the formation and evolution of
cosmic structure in a scenario known as $\Lambda$CDM. In these
cosmological models, gravity is described by General Relativity, dark matter
and dark energy are the major constituents of the Universe (with
normal `baryonic' matter only contributing $\sim 4$ percent), and
density perturbations are seeded by quantum fluctuations in a scalar
field, the inflaton, that dominated the energy density shortly after
the Big Bang. In its most basic (`vanilla') form, the $\Lambda$CDM
model assumes a flat geometry, dark energy is modeled as Einstein's
cosmological constant, neutrino mass is assumed to be negligible, and
the initial power spectrum of density perturbations is assumed to be a
single power-law. Such $\Lambda$CDM cosmologies are described by 5
parameters: the energy densities (in terms of the critical density) of
baryons, $\Omega_\rmb$, and cold dark matter, $\Omega_{\rm dm}$, the
spectral index, $n_\rms$, and normalization, $\sigma_8$, of the
initial power spectrum, and the Hubble parameter, $h = H_0/(100
\kmsmpc)$.  The flat geometry implies that $\Omega_\Lambda = 1 -
\Omega_\rmm = 1 - \Omega_\rmb - \Omega_{\rm dm}$, and is strongly
supported by the location of the first peak in the angular power
spectrum of cosmic microwave background (CMB) temperature fluctuations
(e.g., Balbi \etal 2000; Lange \etal 2001; Pryke \etal 2002;
Netterfield \etal 2002; Ruhl \etal 2003) combined with the results 
on the Hubble constant from
the Hubble Key Project (Freedman \etal 2001).

All these cosmological parameters have now been constrained at the few
($\sim~3$-$20$) percent level by a variety of probes, including, among
others, temperature anisotropies in the CMB (e.g., Spergel \etal 2003,
2007; Reichardt \etal 2009; Dunkley \etal 2009; Komatsu \etal 2009,
2011), Cepheid distances (e.g., Freedman \etal 2001; Sandage \etal
2006; van Leeuwen \etal 2007), high redshift supernovae Ia (e.g.,
Riess \etal 1998; Perlmutter \etal 1999, Astier \etal 2006; Kowalski
\etal 2008), measurements of the primordial deuterium abundance (e.g.,
Burles, Nollett \& Turner 2001; O'Meara \etal 2006), cluster
abundances (e.g., Vikhlinin \etal 2009; Rozo \etal 2010; Benson \etal
2011), cosmic shear (e.g., Benjamin \etal 2007; Fu \etal 2008; Lin
\etal 2011), the integrated Sachs-Wolfe effect (e.g., Ho \etal 2008;
Giannantonio \etal 2008), the Ly$\alpha$ forest (e.g., Viel, Weller \&
Haehnelt 2004; McDonald \etal 2005; Desjacques \& Nusser 2005), and
strong gravitational lensing (e.g., Koopmans \etal 2003; Oguri \etal
2008).  Despite some tension between a few subsets of all these
independent constraints (see e.g., Dunkley \etal 2009 for a
comprehensive overview), overall they are in good mutual agreement,
giving rise to the notion of a true concordance cosmology.

Another potentially powerful probe for cosmology is the (large-scale)
distribution of galaxies. Although stars make a negligible contribution
to the total energy density of the Universe, the light from stars in galaxies
can be observed directly and over cosmological scales, making galaxies
useful tracers of the underlying dark matter density field.
Unfortunately, the connection between galaxies and (dark) matter is
muddled by the fact that galaxies are {\it biased} tracers of the mass
distribution. The main problem is that this `galaxy bias' is known to
be extremely complicated: it is stochastic (e.g., Dekel \& Lahav 1999;
Tegmark \& Bromley 1999), depends on galaxy properties such as
luminosity, color and/or morphological type (e.g., Park \etal 1994;
Guzzo \etal 2000; Norberg \etal 2001, 2002; Zehavi \etal 2005, 2011;
Wang \etal 2007), and is scale dependent on small scales (e.g.,
Pervical \etal 2007; Reid, Spergel \& Bode 2009; Cacciato \etal 2012).
Based on these considerations, it is not surprising that galaxy bias
is generally considered a nuisance when using galaxies to constrain
cosmology. However, galaxy bias also contains a wealth of information
regarding galaxy formation, especially on small scales (e.g., Cacciato
\etal 2012). After all, it is the (poorly understood) physics of
galaxy formation that determines where, how and with what efficiency
galaxies form within the dark matter density field (see Mo, van den
Bosch \& White 2010). Therefore, ideally, one would like to {\it
  simultaneously} solve for cosmology and galaxy bias.

Early attempts to do so used galaxy power spectra measured from the
two-Degree Field Galaxy Redshift Survey (2dFGRS; Colless \etal 2003)
and/or Sloan Digital Sky Survey (SDSS; York \etal 2000) combined with
a phenomenological fitting function for the non-linear,
scale-dependent bias and marginalized over its free parameters (e.g.,
Cole \etal 2005; Tegmark \etal 2006; Padmanabhan \etal 2007). However,
this approach has two problems.  First, the fitting function used to
describe the galaxy bias has no logical connection to galaxy
formation. For this reason the free parameters used to parameterize
the galaxy bias are merely considered as nuisance parameters.  Second,
as pointed out by S\'anchez \& Cole (2008) and Yoo \etal (2009), the
particular fitting function used is often a poor description of the
true scale dependence of galaxy bias, causing biased estimates of the
cosmological parameters, especially for non-vanilla $\Lambda$CDM
cosmologies (e.g., Hamann \etal 2008). It is generally believed that
the latter problem explains why there is still some tension between
cosmological parameters (mainly $\Omega_\rmm$) inferred from different
galaxy power spectra, or from using the same power spectrum, but using
data covering different scales (e.g., Percival \etal 2007; S\'anchez
\& Cole 2008; Dunkley \etal 2009)

This indicates that it is prudent to use a realistic, physically
motivated model for galaxy bias. Under the assumption that all
galaxies reside in dark matter haloes, a natural, realistic model for
galaxy bias is provided by halo occupation models, which describe, in
a statistical sense, how galaxies are distributed over dark matter
haloes (e.g., Jing, Mo \& B\"orner 1998; Peacock \& Smith 2000;
Scoccimarro \etal 2001; Berlind \& Weinberg 2002; Yang, Mo \& van den
Bosch 2003). When combined with the halo model, which describes the
(non-linear) matter distribution in terms of the dark matter halo
building blocks (e.g., Neyman \& Scott 1952; Seljak 2000; Ma \& Fry
2000; Cooray \& Sheth 2002), these halo occupation models provide a
complete, accurate, and easy-to-interpret description of galaxy bias,
all the way from the large, linear scales down to the small,
non-linear scales of individual dark matter haloes.

The problem with this approach, though, is that the halo occupation
models are also cosmology dependent, so that one typically needs
constraints in addition to just large scale clustering. Several
attempts have been made along these lines.  Abazajian \etal (2005)
simply used strong priors from the WMAP results.  Yang \etal (2004)
and Tinker \etal (2007) used peculiar velocities as inferred from the
redshift space distortions in the two-point correlation function, and
argued for a relatively low value of $\sigma_8 \sim 0.7 - 0.75$ (for a
vanilla $\Lambda$CDM cosmology with $\Omega_\rmm = 0.3$). Similar
conclusions were reached by van den Bosch \etal (2003b) and Tinker
\etal (2005), who used constraints on the (average) mass-to-light
ratios of clusters, rather than peculiar velocities.  Somewhat
puzzling, a more recent analysis by Tinker \etal (2012) used the
mass-to-number ratio of clusters, rather than the mass-to-light
ratios, and found a relatively {\it high} value for $\sigma_8$ of
$0.85 \pm 0.05$ (again for $\Omega_\rmm = 0.3$).

In this paper we use a combination of galaxy clustering and
galaxy-galaxy lensing, as well as constraints on galaxy abundances, in
order to simultaneously constrain cosmology and galaxy bias.  Since
galaxy-galaxy lensing probes the mass associated with the lensing
galaxies, this is similar to using mass-to-light ratios as
constraints. It has the advantage, though, that it probes
mass-to-light ratios over a wide range in halo masses, and that the
same halo occupation model used to compute the clustering of galaxies
can also be used to compute the galaxy-galaxy lensing signal (e.g.,
Guzik \& Seljak 2002; Yoo \etal 2006; Cacciato \etal 2009).  A first
application of this idea by Seljak \etal (2005), when combined with
WMAP constraints, yielded a relative high value for $\sigma_8$ of
$0.88 \pm 0.06$ (for $\Omega_\rmm \simeq 0.25$). More recently, two
different analyses based on the same galaxy-galaxy lensing data by
Cacciato \etal (2009) and Li \etal (2009) both argued that a flat
$\Lambda$CDM cosmology with $(\Omega_\rmm,\sigma_8) = (0.238, 0.734)$
is in much better agreement with the data than a $(0.3,0.9)$ model,
thus favoring again a relatively low value for $\sigma_8$.  In this
paper we improve on all these previous methods by (i) simultaneously
constraining galaxy bias and cosmology, (ii) using a much more
accurate analytical model, (iii) using the latest clustering data,
(iv) modelling the latter accounting for residual redshift space distortions,
(v) marginalizing over uncertainties related to the detailed structure
of dark matter haloes and the scale dependence of the halo bias, and
(vi) subjecting the analysis to a number of detailed tests that address
how the results depend on certain assumptions inherent to the model.

This paper is the third in a series. In van den Bosch \etal (2012;
hereafter Paper~I), we presented the analytical model, which we
calibrated and tested using detailed mock catalogs constructed from
high resolution $N$-body simulations. We demonstrated that our
analytical model is accurate at the level of 10\% or better over the
entire range of scales covered by the data. In More \etal (2012b;
hereafter Paper~II), we presented a Fisher matrix analysis to identify
parameter-degeneracies and to assess the accuracy with which various
cosmological parameters can be constrained using our methodology.  We
demonstrated that the method can simultaneously constrain halo
occupation statistics and cosmology, and we forecasted that, using
existing data from the SDSS, we should be able to put constraints on
$\Omega_\rmm$ and $\sigma_8$ that are among the tightest ever
achieved. In this paper we apply our method to existing data from the
SDSS\footnote{A preliminary version of the main results presented in
  this paper were published in a conference proceedings by More \etal
  (2012a).}. Although, as demonstrated in paper~II, our method is also
able to constrain extensions to the vanilla $\Lambda$CDM cosmology,
such as neutrino mass and the equation of state of dark energy, in
this paper we focus solely on vanilla $\Lambda$CDM cosmologies, and in
particular on constraining the combination
$(\Omega_\rmm,\sigma_8)$. We defer constraining neutrino mass, dark
energy, and other modifications to the vanilla $\Lambda$CDM cosmology
to future papers.

This paper is organized as follows. In \S\ref{sec:data}, we introduce
the SDSS data used to constrain our models. In \S\ref{sec:model}, we
briefly review our analytical model to compute the galaxy luminosity
function, the galaxy-galaxy correlation function, and the
galaxy-galaxy lensing signal using the halo model combined with a
model that describes halo occupation statistics as function of galaxy
luminosity. 
The Bayesian analysis,
used to infer posterior distributions for the cosmological parameters
and for the parameters that describe the halo occupation
statistics, is described in \S\ref{sec:method}. 
Our main results are presented in \S\ref{sec:results},
while \S\ref{sec:systematics} describes a number of tests that address
the sensitivity of our results to several model assumptions.  We
summarize our findings in \S\ref{sec:conclusions}.

Throughout this paper, unless specifically stated otherwise, all radii
and densities are in comoving units, and log is used to refer to the
10-based logarithm. Quantities that depend on the Hubble parameter are
written in units of $h$, defined above.


\section{Data}
\label{sec:data}

The data used to constrain our models consists of three
components: galaxy abundances, in the form of the galaxy luminosity
function, galaxy clustering, in the form of projected correlation
functions for six different luminosity bins, and galaxy-galaxy
lensing, in the form of excess surface densities (ESD), once again for
six different luminosity bins. All these measurements are obtained
from the Sloan Digital Sky Survey (SDSS; York \etal 2000).

We use the $r$-band galaxy luminosity function, $\Phi(L)$, (hereafter
LF) of Blanton \etal (2003a)\footnote{available at
  http://cosmo.nyu.edu/mb144/lf.html}, sampled at 32 magnitudes in the
range $-23.0 \leq {^{0.1}M}_r-5\log h \leq -18.0$, where ${^{0.1}M}_r$
indicates the $r$-band magnitude of galaxies K$+$E corrected to
$z=0.1$ following the procedure of Blanton \etal (2003b). For each of
these magnitude bins we use the (statistical) errors on $\Phi$ quoted
by Blanton \etal (2003a). Unfortunately, we do not have a full
covariance matrix for this data set. There are two main sources of
covariance for the luminosity function: magnitude errors, which cause
covariance between neighboring magnitude bins, and large scale
structure (`sample variance')\footnote{Following Scott, Srednicki \&
  White (1994) we use the term `sample variance', rather than the more
  common `cosmic variance'}.  The LF of Blanton \etal is sampled at
magnitude intervals of $\delta m_r = 0.00986$ mag, which is much
smaller than the typical error on individual magnitudes ($\sim 0.05$
mag).  In order to suppress the covariance due to these magnitude
errors, we only sample the LF of Blanton et al. at 32 magnitude
intervals of $\delta m_r = 0.1578$, roughly three times as large as
the typical magnitude error.  This should eliminate virtually all
covariance due to errors in the magnitudes of individual galaxies.
This leaves the covariance due to sample variance, which can be
`modeled' to reasonable accuracy as an up or down shift of the entire
LF (i.e.,. the error bars are fully covariant, e.g. Blanton et al. 2003a).  
Throughout this study, we ignore this sample variance, which effectively implies that we
assume that the volume probed by the SDSS is a fair representation of
the average Universe (see discussion in \S\ref{sec:variance}).

The galaxy clustering data used in this paper are taken from Zehavi
\etal (2011), and based on the SDSS DR7 (Abazajian \etal 2009).  Using
all galaxies in the main galaxy sample with apparent magnitudes $14.5
< m_r < 17.6$, Zehavi \etal measured the projected correlation
functions, $w_\rmp(r_\rmp)$, over the radial range $0.2 h^{-1}\Mpc <
r_\rmp < 20 h^{-1}\Mpc$ for six volume limited samples (see Table~1).
These have been obtained according to
\begin{equation}\label{wpzspace}
w_\rmp(r_{\rm p}) = 2 \int_{0}^{r_{\rm max}} \xi_{\rm gg}(r_{\rm p},r_{\pi}) 
\,{\rm d}r_{\pi}\,.
\end{equation}
with $r_{\rm max} = 40 h^{-1} \Mpc$ or $60 h^{-1}\Mpc$, depending on
the luminosity sample used (see Table~1).  Here $r_{\rm p}$ is the
projected separation between two galaxies, $r_{\pi}$ is the
redshift-space separation along the line-of-sight, and $\xi_{\rm
  gg}(r_{\rm p},r_{\pi})$ is the two-dimensional correlation function,
which is anisotropic due to the presence of peculiar velocities. As
discussed in Paper~I, the fact that $r_{\rm max}$ is finite results in
residual redshift space distortions that need to be corrected for in
the modeling (see also \S~\ref{sec:noRSD}).  The errors on
$w_\rmp(r_\rmp)$ are characterized by their full covariance matrices,
kindly provided to us in electronic format by I. Zehavi, and obtained
from 144 spatially contiguous subsamples using the jackknife
technique.

Finally, for the galaxy-galaxy lensing data we use the excess surface
densities, $\Delta\Sigma(R)$, covering the radial range $0.04
h^{-1}\Mpc < R < 2 h^{-1}\Mpc$, obtained by Seljak \etal (2005) and
Mandelbaum \etal (2006), and kindly provided to us in electronic
format by R. Mandelbaum. These measurements have been obtained using a
catalogue of $351,507$ lens galaxies with apparent magnitude $m_r <
17.77$ in the redshift range $0.02 < z < 0.35$ taken from the main
galaxy catalogue of the SDSS Data Release 4 (Adelman-McCarthy \etal
2006).  This sample is split in 8 (flux-limited) luminosity
bins. However, since both the faintest and brightest bins have
extremely poor signal-to-noise, we only use the 6 intermediate
luminosity bins listed in Table~2.  We refer the reader to Mandelbaum
\etal (2006) for a detailed description of the data and of the method
used to determine the excess surface density profiles.  Since the
covariance in $\Delta\Sigma(R)$ is only very small over the radial
scales covered by the data (R. Mandelbaum, private communication), we
only use the diagonal elements in what follows.
\begin{table}\label{tab:wpdata}
\caption{Properties of the projected correlation function data}
\begin{center}
\begin{tabular}{ccccrc}
\hline\hline
  $^{0.1}M_r - 5 \log h$ & $z_{\rm min}$ & $z_{\rm max}$ & $\langle z \rangle$ 
& $N_{\rm gal}$ & $r_{\rm max}$ \\
  (1) & (2) & (3) & (4) & (5) & (6) \\
\hline
 $(-18,-17]$ & 0.011 & 0.026 & 0.021 &   5,965 & 40 \\
 $(-19,-18]$ & 0.017 & 0.042 & 0.032 &  18,200 & 40 \\
 $(-20,-19]$ & 0.027 & 0.064 & 0.050 &  44,348 & 60 \\
 $(-21,-20]$ & 0.042 & 0.106 & 0.082 & 108,629 & 60 \\
 $(-22,-21]$ & 0.066 & 0.159 & 0.123 &  73,746 & 60 \\
 $(-23,-22]$ & 0.103 & 0.245 & 0.187 &  10,251 & 60 \\
\hline\hline
\end{tabular}
\end{center}
\medskip
\begin{minipage}{\hssize}
  The volume-limited galaxy samples used by Zehavi \etal (2011) to
  measure the projected correlation function, $w_\rmp(r_\rmp)$.  For
  each of these samples column (1) lists the magnitude range in
  ${^{0.1}M}_r-5\log h$, columns~(2), (3) and (4) list the minimum,
  maximum and mean redshifts, respectively, column~(5) lists the
  number of galaxies, and column (6) lists the value of $r_{\rm max}$
  (in $h^{-1}\Mpc$), defined in Eq.~(\ref{wpzspace}).
\end{minipage}
\end{table}
Throughout this paper we refer to these three data sets as the LF, WP
and ESD data. Note that each of these data sets adopted a flat
$\Lambda$CDM cosmology with $\Omega_\rmm=0.3$ (LF and WP) or
$\Omega_\rmm = 0.27$ (ESD) when computing distances and/or absolute
magnitudes. Changing the assumed cosmology, as we do in our analysis,
in principle therefore also has a small impact on the observational
measurements by changing the distance-redshift relation and thus
shifting galaxies among luminosity bins and galaxy pairs among radial
separation bins. However, in this paper we restrict ourselves to
cosmologies that only differ mildly from a $\Lambda$CDM cosmology with
$\Omega_\rmm=0.3$. Even at our outer redshift limit of z = 0.2, the
effect of lowering $\Omega_\rmm$ from 0.3 to 0.25 is only 1 percent in
distance, so the measurements used here are effectively independent of
cosmological parameters within their observational
uncertainties. Therefore, we ignore this small effect in what follows,
and always use the observational data as described above.


\section{Model Description} 
\label{sec:model}
\begin{table}\label{tab:esddata}
  \caption{Properties of the excess surface density data}
\begin{center}
\begin{tabular}{ccr}
\hline\hline
  $^{0.1}M_r - 5 \log h$ & $\langle z \rangle$ & $N_{\rm lens}$ \\
  (1) & (2) & (3) \\
\hline
 $(-19.0,-18.0]$ & 0.047 &  29,730 \\
 $(-20.0,-19.0]$ & 0.071 &  85,766 \\
 $(-21.0,-20.0]$ & 0.10  & 141,976 \\
 $(-21.5,-21.0]$ & 0.14  &  60,994 \\
 $(-22.0,-21.5]$ & 0.17  &  34,920 \\
 $(-22.5,-22.0]$ & 0.20  &  13,067 \\
\hline\hline
\end{tabular}
\end{center}
\medskip
\begin{minipage}{\hssize}
  The flux-limited galaxy samples used by Mandelbaum \etal (2006) to
  measure the excess surface density profiles, $\Delta\Sigma(R)$.
  For each of these samples column (1) lists the magnitude range in
  ${^{0.1}M}_r-5\log h$, column~(2) lists the mean redshift, and
  column (3) lists the number of lens galaxies used. 
\end{minipage}
\end{table}

As described in detail in Paper~I, the observables\footnote{The
  projected correlation function and the excess surface density are
  modeled per luminosity bin $[L_1,L_2]$. However, to keep the
  notation concise, we do not explicitly write down the dependence on
  $L_1$ and $L_2$.}  $\Phi(L,z)$, $w_\rmp(r_\rmp,z)$ and $\Delta
\Sigma(R,z)$ can be computed for a given cosmology, which determines
the properties of the dark matter distribution (e.g, halo mass
function, halo bias function, halo density profiles), and a given
description of the galaxy-dark matter connection (i.e., halo
occupation statistics). The dependence on redshift is included to
emphasize that we model each observable at its mean redshift. For the
luminosity function we adopt $z = 0.1$, while the mean redshifts for
the different luminosity bins of the $w_\rmp(r_\rmp)$ and
$\Delta\Sigma(R)$ measurements are listed in Tables~1 and~2,
respectively.

In this section we give a concise overview of our model. Readers
interested in a more thorough description are referred to Paper~I,
while those readers that are already familiar with our model, or that
are mainly interested in the results, may want to skip this section
and proceed immediately to \S~\ref{sec:method}.

\subsection{Cosmological Parameters}
\label{sec:cosmopar}

Throughout this paper we consider `vanilla' $\Lambda$CDM cosmologies
in which gravity is described by standard General Relativity, neutrino
mass is negligible, the initial power spectrum is a single power-law,
and dark energy is modeled as Einstein's cosmological constant with
$\Omega_{\Lambda} = 1 - \Omega_\rmm$. These cosmologies are completely
specified by five parameters; the $z=0$ matter density in units
of the critical density, $\Omega_\rmm$,
the normalization of the matter power spectrum, $\sigma_8$, the Hubble
parameter $h = H_0/(100 \kmsmpc)$, the initial spectral index of the
matter power spectrum, $n_\rms$, and the $z=0$ baryon density in units
of the critical density, $\Omega_{\rm b}$. Hence, our cosmological model parameters are described by the vector
\begin{equation}\label{lambdaCOSMO}
{\bolds\lambda}^{\rm cosmo} \equiv (\Omega_\rmm, \sigma_8, h, n_\rms, \Ombh)\,.
\end{equation}
Note that the baryon density only enters in our analysis in the
transfer function, and always in the combination $\Ombh$.

The main goal of this paper is to use the observational data on
$\Phi(L)$, $w_\rmp(r_\rmp)$ and $\Delta\Sigma(R)$ discussed in
\S\ref{sec:data} to constrain $\Omega_\rmm$ and $\sigma_8$. Throughout
this paper we will therefore not use any priors on these two
parameters. For $h$, $n_\rms$ and $\Omega_\rmb$, on the other hand, we
include prior information from the seven year analysis of the cosmic
microwave background data from WMAP (hereafter WMAP7; Komatsu \etal
2011), as described in \S\ref{sec:method} below.  For this reason we
refer to $h$, $n_\rms$ and $\Omega_\rmb$ as our secondary cosmological
parameters in what follows.

\subsection{The Conditional Luminosity Function}
\label{sec:clf}

Under the assumption that each galaxy resides in a dark matter halo,
$\Phi(L,z)$, $w_\rmp(r_\rmp,z)$ and $\Delta\Sigma(R,z)$ can be
computed using a statistical description of how galaxies are
distributed over dark matter haloes of different mass. To that extent
we use the conditional luminosity function (hereafter CLF) introduced
by Yang \etal (2003).  The CLF, $\Phi(L|M) \rmd L$, specifies the {\it
  average} number of galaxies with luminosities in the range $L \pm
\rmd L/2$ that reside in a halo of mass $M$.  We split the CLF in two
components,
\begin{equation}
\Phi(L|M) = \Phi_\rmc(L|M) + \Phi_\rms(L|M)\,,
\end{equation}
where $\Phi_\rmc(L|M)$ describes the contribution due to central
galaxies (defined as those galaxies that reside at the center of their
host halo), while $\Phi_\rms(L|M)$ characterizes satellite galaxies
(those that orbit around a central).  Throughout we ignore a potential
redshift dependence of the CLF. Since the data that we use to
constrain the CLF only covers a narrow range in redshift, this
assumption will not have a strong impact on our results.

Our parameterization of the CLF model is motivated by the results
obtained by Yang, Mo \& van den Bosch (2008) from a large galaxy group
catalogue (Yang \etal 2007) extracted from the SDSS Data Release 4,
and by Tal \etal (2012) from a study of the luminosity function of
satellite galaxies of luminous red galaxies.  In particular, the CLF
of central galaxies is modeled as a log-normal function:
\begin{equation}\label{phi_c}
\Phi_\rmc(L|M) \,{\rmd}L = {\log\, e \over {\sqrt{2\pi} \, \sigma_\rmc}} 
{\rm exp}\left[- { {(\log L  -\log L_\rmc )^2 } \over 2\,\sigma_\rmc^2} \right]\,
{\rmd L \over L}\,,
\end{equation}
and the satellite term as a modified Schechter function:
\begin{equation}\label{phi_s}
\Phi_\rms(L|M)\,{\rmd}L = \phi^*_\rms \,
\left({L\over L^*_\rms}\right)^{\alpha_\rms + 1} \,
{\rm exp} \left[- \left ({L\over L^*_\rms}\right )^2 \right] {\rmd L \over L}\,,
\end{equation}
which decreases faster than a Schechter function at the bright end.
Note that $L_\rmc$, $\sigma_\rmc$, $\phi^*_\rms$, $\alpha_\rms$ and
$L^*_\rms$ are all functions of the halo mass $M$.

Following Cacciato \etal (2009), and motivated by the results of Yang
\etal (2008) and More \etal (2009, 2011), we assume that
$\sigma_\rmc$, which expresses the scatter in $\log L$ of central
galaxies at fixed halo mass, is a constant (i.e.  is independent of
halo mass and redshift).  In addition, for $L_\rmc$, we adopt the
following parameterization;
\begin{equation}\label{LcM}
L_\rmc(M) = L_0 {(M/M_1)^{\gamma_1} \over 
\left[1 + (M/M_1) \right]^{\gamma_1-\gamma_2}}\,.
\end{equation}
Hence, $L_\rmc \propto M^{\gamma_1}$ for $M \ll M_1$ and $L_c \propto
M^{\gamma_2}$ for  $M \gg  M_1$. Here $M_1$  is a  characteristic mass
scale, and $L_0 =  2^{\gamma_1-\gamma_2} L_c(M_1)$ is a normalization. 

For the satellite galaxies we adopt
\begin{equation}
L^*_\rms(M)  = 0.562 L_\rmc(M)\,,
\end{equation}
\begin{equation}\label{alpha}
\alpha_\rms(M) = \alpha_\rms
\end{equation}
(i.e., the faint-end slope of $\Phi_\rms(L|M)$ is independent of mass
and redshift), and
\begin{equation}\label{phi}
\log[\phi^*_\rms(M)] = b_0 + b_1 (\log M_{12}) + b_2 (\log M_{12})^2\,,
\end{equation}
with $M_{12}=M/(10^{12} h^{-1}\Msun)$. Note that neither of these
functional forms has a physical motivation; they merely were found to
adequately describe the results obtained by Yang \etal (2008) from the
SDSS galaxy group catalog.

To summarize, our parameterization of the CLF thus has a total of nine
free parameters, characterized by the vector
\begin{equation}\label{lambdaCLF}
{\bolds\lambda}^{\rm CLF} \equiv (\log M_1, \log L_{0}, \gamma_1, 
\gamma_2, \sigma_\rmc, \alpha_\rms, b_0, b_1, b_2) \, . 
\end{equation}

\subsection{Galaxy Luminosity Function}

Once the CLF is specified, the galaxy luminosity function at
redshift $z$, $\Phi(L,z)$, simply follows from integrating over the
halo mass function, $n(M,z)$;
\begin{equation}
\Phi(L,z) = \int \Phi(L|M) \, n(M,z) \, \rmd M\,.
\end{equation}
In what follows, we will always be concerned with galaxies in a
specific luminosity interval $[L_1,L_2]$. The average number density
of such galaxies follows from the CLF according to
\begin{equation}\label{avgngal}
\bar{n}_\rmg(z) = \int \langle N_\rmg|M \rangle \, n(M,z) \, \rmd M\,,
\end{equation}
where
\begin{equation}\label{avgnm}
\langle N_\rmg|M\rangle = \int_{L_1}^{L_2} \Phi(L|M) \rmd L\,,
\end{equation}
is the average number of galaxies with $L_1 < L < L_2$ that reside in
a halo of mass $M$.

\subsection{Power-Spectra}
\label{sec:powerspectra}

The first step towards computing the projected correlation functions,
$w_\rmp(r_\rmp,z)$, and ESD profiles, $\Delta\Sigma(R,z)$, is to
compute the galaxy-galaxy power spectrum, $P_{\rm gg}(k,z)$, and the
galaxy-matter cross power spectrum, $P_{\rm gm}(k,z)$.  These power spectra
are the Fourier space analogs of the galaxy-galaxy and galaxy-matter
correlation functions.

The galaxy-galaxy power spectrum can be expressed as a sum of the
one-halo (1h) and the two-halo (2h) terms, each of which can be
further subdivided based upon the type of galaxies (central or
satellite) that contribute to the power spectrum, i.e.,
\begin{eqnarray}
P_{\rm gg}(k) & = & 2 P^{\rm 1h}_{\rm cs}(k) + P^{\rm 1h}_{\rm ss}(k) \nonumber \\ 
& + & P^{\rm 2h}_{\rm cc}(k) + 2 P^{\rm 2h}_{\rm cs}(k) + P^{\rm 2h}_{\rm ss}(k)\,.
\end{eqnarray}
Similarly, the galaxy-matter power spectrum can be written as
\begin{equation}
P_{\rm gm}(k) = P^{\rm 1h}_{\rm cm}(k) + P^{\rm 1h}_{\rm sm}(k) 
+ P^{\rm 2h}_{\rm cm}(k) + P^{\rm 2h}_{\rm sm}(k)\,.
\end{equation} 
As shown in paper~I, these terms can be written in compact form as
\begin{equation}\label{P1h}
P^{\rm 1h}_{\rm xy}(k,z) = \int \calH_\rmx(k,M,z) \, \calH_\rmy(k,M,z) \, 
n(M,z) \, \rmd M,
\end{equation}
\begin{eqnarray}\label{P2h}
\lefteqn{P^{\rm 2h}_{\rmx\rmy}(k,z) =
\int \rmd M_1 \, \calH_\rmx(k,M_1,z) \, n(M_1,z) } \nonumber \\
& & \int \rmd M_2 \, \calH_\rmy(k,M_2,z) \, n(M_2,z) \,
Q(k|M_1,M_2,z)\,,
\end{eqnarray}
where `x' and `y' are either `c' (for central), `s' (for satellite),
or `m' (for matter), $Q(k|M_1,M_2,z)$ describes the power-spectrum of
haloes of mass $M_1$ and $M_2$ (see Appendix~\ref{app:Qk}), and we
have defined
\begin{equation}\label{calHm}
\calH_\rmm(k,M,z) = {M \over \bar{\rho}_{\rmm}} \, \tilde{u}_\rmh(k|M,z)\,,
\end{equation}
\begin{equation}\label{calHc}
\calH_\rmc(k,M,z) = \calH_\rmc(M,z) = 
{\langle N_\rmc|M \rangle \over \bar{n}_{\rmg}(z)} \,,
\end{equation}
and
\begin{equation}\label{calHs}
\calH_\rms(k,M,z) = {\langle N_\rms|M \rangle \over \bar{n}_{\rmg}(z)} \,  
\tilde{u}_\rms(k|M,z)\,.
\end{equation}
Here $\langle N_\rmc|M \rangle$ and $\langle N_\rms|M \rangle$ are the
average number of central and satellite galaxies in a halo of mass
$M$, which follow from Eq.~(\ref{avgnm}) upon replacing $\Phi(L|M)$ by
$\Phi_\rmc(L|M)$ and $\Phi_\rms(L|M)$, respectively.  Furthermore,
$\tilde{u}_\rms(k|M)$ is the Fourier transform of the normalized
number density distribution of satellite galaxies that reside in a
halo of mass $M$, and $\tilde{u}_\rmh(k|M)$ is the Fourier transform
of the normalized density distribution of dark matter within a halo of
mass $M$.

\subsection{Computing $w_\rmp(r_\rmp,z)$ and $\Delta\Sigma(R,z)$}
\label{sec:wpesd}

Once $P_{\rm gg}(k,z)$ and $P_{\rm gm}(k,z)$ have been determined, it
is fairly straightforward to compute the projected galaxy-galaxy
correlation function, $w_\rmp(r_\rmp,z)$, and the excess surface
density (ESD) profile, $\Delta\Sigma(R,z)$. We start by Fourier
transforming the power-spectra to obtain the two-point correlation
functions:
\begin{eqnarray}\label{xiFTfromPK}
\xi_{\rm gx}(r,z) = {1 \over 2 \pi^2} \int_0^{\infty} P_{\rm gx}(k,z) 
{\sin kr \over kr} \, k^2 \, \rmd k\,, 
\end{eqnarray}
where `x' is either `g' (for galaxies) or `m' (for matter).

The projected galaxy-galaxy correlation function, $w_\rmp(R,z)$, is
related to the real-space galaxy-galaxy correlation function,
$\xi_{\rm gg}(r,z)$, according to
\begin{eqnarray}\label{wpRRSD}
w_\rmp(r_\rmp,z) & = & 2 \int_{0}^{r_{\rm max}} \xi_{\rm gg}(r_\rmp,r_{\pi},z) 
\, \rmd r_{\pi} \nonumber \\
& = &  2 \sum_{l=0}^2 \int_0^{r_{\rm max}} \xi_{2l}(s,z)\,\calP_{2l}(r_{\pi}/s)
\,\rmd r_{\pi} \,.
\end{eqnarray}
Here $r_{\rm max}$ is the maximum integration range used for the data
(see Table~1), $r_{\pi}$ is the separation perpendicular to the
line-of-sight, $s = \sqrt{r^2_\rmp + r^2_{\pi}}$ is the separation
between the galaxies in redshift space, $\calP_l(x)$ is the $l^{\rm
  th}$ Legendre polynomial, and $\xi_0$, $\xi_2$, and $\xi_4$ are
given by
\begin{equation}\label{monopole}
\xi_0(r,z) = \left( 1 + {2 \over 3}\beta + {1 \over 5}\beta^2\right) \,
\xi_{\rm gg}(r,z)\,,
\end{equation}
\begin{equation}\label{quadrupole}
\xi_2(r,z) = \left( {4 \over 3}\beta + {4 \over 7}\beta^2\right) \,
\left[\xi_{\rm gg}(r,z) - 3 J_3(r,z)\right]\,,
\end{equation}
\begin{equation}\label{octopole}
\xi_4(r,z) = {8 \over 35}\beta^2 \, \left[\xi_{\rm gg}(r,z) + 
{15\over 2} J_3(r,z) - {35\over 2}J_5(r,z) \right]\,,
\end{equation}
where
\begin{equation}\label{Jintegral}
J_n(r,z) = {1 \over r^n} \int_0^r \xi_{\rm gg}(y,z) \, y^{n-1} \, \rmd y\,. 
\end{equation}
and
\begin{equation}\label{betapar}
\beta = \beta(z) = {1 \over \bar{b}(z)} 
\left({\rmd {\rm ln} D \over \rmd {\rm ln} a}\right)_z
\simeq {\Omega^{0.6}_\rmm(z) \over \bar{b}(z)}
\end{equation}
with $a = 1/(1+z)$ the scale factor, $D(z)$ the linear growth
rate, and 
\begin{equation}\label{avbias}
\bar{b}(z) = {1 \over \bar{n}_\rmg(z)} \int \langle N_\rmg|M \rangle \,
b_\rmh(M,z) \, n(M,z) \, \rmd M\,,
\end{equation}
the mean bias of the galaxies in consideration. Note that
Eq.~(\ref{wpRRSD}) accounts for the large scale redshift space
distortions due to infall (the `Kaiser'-effect), which is necessary
because the measurements for $w_\rmp(\rmp)$ have been obtained using a
finite $r_{\rm max}$.  Note that the $\xi_{\rm gg}(r)$ in
Eqs.~(\ref{monopole})-(\ref{octopole}) is the non-linear galaxy-galaxy
correlation function.  Although the Kaiser formalism (Kaiser 1987) is
only strictly valid in the linear regime, this simple modification
results in a more accurate correction for the residual redshift space
distortions (see Paper~I for details). As shown in Paper~I, not taking
these residual redshift space distortions into account results in
systematic errors that can easily exceed 20 percent on large scales
($r_\rmp \gta 10 h^{-1}\Mpc$), causing systematic errors in the
inferred galaxy bias (see also Padmanabhan, White \& Eisenstein 2007;
Norberg \etal 2009; Baldauf \etal 2010; More 2011).

Finally, the excess surface density profile, $\Delta \Sigma(R,z)$, is
defined as
\begin{equation}\label{shear} 
\Delta\Sigma(R,z) = 
{2\over R^2} \int_0^R \Sigma(R',z) \, R' \, \rmd R'  - \Sigma(R,z) 
\end{equation}
Here $\Sigma(R,z)$ is the projected surface mass density, which is
related to the galaxy-dark matter cross correlation, $\xi_{\rm
  gm}(r,z)$, according to
\begin{equation}\label{Sigma_approx}
\Sigma(R,z) = 2\bar{\rho}_\rmm \int_{R}^{\infty} 
\left[1+\xi_{\rm gm}(r,z)\right] \, { r \rmd r \over \sqrt{r^2 - R^2}}\,.
\end{equation}

\subsection{Model Ingredients}
\label{sec:ingredients}

In this subsection we briefly describe the ingredients of our model.
A more detailed description can be found in Paper~I.
\begin{table*}\label{tab:modelpar}
\caption{List of models}
\begin{tabular}{cccccccccccc}
\hline\hline
Model & $\psi$ & $\eta$ & $\calR$ & $\calP$ & RRSD & SGW & $\chi^2_{\rm LF}$ & $\chi^2_{\rm Wp}$ & $\chi^2_{\rm ESD}$ & $\chi^2_{\rm prior}$ & $\chi^2_{\rm red}$ \\
  (1) & (2) & (3) & (4) & (5) & (6) & (7) & (8) & (9) & (10) & (11) & (12) \\
\hline
\modF0  & $\calG(0.9,0.15)$ & $\calG(0.0,0.1)$ & $1.0$             & $1.0$            & yes & yes & 47 & 236 & 117 &  4 & 2.36 \\
\modG0  & $\calG(0.9,0.15)$ & $\calG(0.0,0.1)$ & $\calG(1.0,0.25)$ & $1.0$            & yes & yes & 40 & 240 & 118 &  5 & 2.36 \\
\modD10 & $\calG(0.9,0.15)$ & $\calG(0.0,0.1)$ & $1.0$             & $\calG(1.0,0.1)$ & yes & yes & 52 & 229 & 113 &  7 & 2.35 \\
\modJ0b & $\calG(0.9,0.15)$ & $\calG(0.0,0.1)$ & $1.0$             & $1.0$            & no  & yes & 48 & 218 & 114 &  5 & 2.25 \\
\modK0  & $\calG(0.9,0.15)$ & $\calG(0.0,0.1)$ & $1.0$             & $1.0$            & yes & no  & 32 & 155 & 111 & 16 & 1.99 \\
\hline\hline
\end{tabular}
\medskip
\begin{minipage}{\hdsize}
  List of models described in the text. Column (1) indicates the model
  name. Columns (2) to (5) indicate the priors used for the parameters
  $\psi$, $\eta$, $\calR$ and $\calA_\rmP$.  Here
  $\calG(\bar{x},\sigma)$ indicates a Gaussian prior centered on
  $\bar{x}$ and with a standard deviation $\sigma$.  Column (6)
  indicates whether or not the model corrected for residual redshift
  space distortions (RRSD), and Column (7) indicates whether or not
  the Sloan great wall (SGW) is included in the clustering data.
  Finally, Columns (8) to (12) list the various $\chi^2$ values of the
  best-fit models in each MCMC chain. The reduced $\chi^2$ value in
  Column (12) is computed for $N_{\rm dof}= 171$ degrees of freedom
  (see \S\ref{sec:results}).
\end{minipage}
\end{table*}

In our fiducial model we require both the linear and the non-linear
power spectra of the matter distribution, $P^{\rm lin}_{\rm mm}(k,z)$
and $P_{\rm mm}(k,z)$, respectively. The latter enters in the
computation of the halo-halo correlation function, as described in
detail in Appendix~\ref{app:Qk}. Throughout we compute $P_{\rm
  mm}(k,z)$ using the fitting formula of Smith \etal
(2003)\footnote{We use the small modification suggested on John
  Peacock's website http://www.roe.ac.uk/$\sim$jap/haloes/, although
  we have verified that this has no significant impact on any of our
  results.}, while for $P^{\rm lin}_{\rm mm}(k,z)$ we use the linear
transfer function of Eisenstein \& Hu (1998), neglecting any
contribution from neutrinos and assuming a CMB temperature of 2.725K
(Mather \etal 1999).

Throughout, we define dark matter haloes as spheres with an average
density of 200 times the background density. We assume that their
density profiles follow the NFW profile (Navarro, Frenk \& White
1997), with a concentration-mass relation given by
\begin{equation}\label{concmass}
c(M,z) = (1 + \eta) \, \bar{c}(M,z)\,.
\end{equation}
Here $\bar{c}(M,z)$ is the average concentration-mass relation of
Macci\`o \etal (2007), properly converted to our definition of halo
mass. We treat $\eta$ as a free nuisance parameter which accounts for
(i) the fact that there is an uncertainty of $\sim 10$ percent in the
average concentration mass relation as obtained by different authors
(e.g., Navarro, Frenk \& White 1997; Eke, Navarro \& Steinmetz 2001;
Bullock \etal 2001; Macci\'o \etal 2007; Zhao \etal 2009), and (ii)
realistic dark matter haloes are triaxial, rather than spherical, have
substructure, and have scatter in the concentration-mass relation. As
shown in Paper~I, setting $\eta = \pm 0.2$ modifies the one-halo term
of $\xi_{\rm gm}(r)$ by more than 20 percent on small scales ($r < 0.1
h^{-1}\Mpc$). As discussed at length in Paper~I, an uncertainty in
$\eta$ of this amount is more than adequate to capture the
inaccuracies in our model that arise from the various
oversimplifications and uncertainties regarding the structure of dark
matter haloes. Hence, in what follows, we adopt a Gaussian prior on
$\eta$, centered on $\eta = 0$ and with a standard deviation
$\sigma_\eta = 0.1$.

For the halo mass function, $n(M,z)$, and the halo bias function,
$b_\rmh(M,z)$, we use the fitting functions of Tinker \etal
(2010). Note that these functions obey the normalization condition
\begin{equation}\label{eq:integralconstraint}
{1\over\overline{\rho}(z)} \int M \, n(M,z) \, b_\rmh(M,z) \, \rmd M = 1
\end{equation}
as required by the fact that, on average, on large scales matter is unbiased
with respect to itself. As described in Appendix~\ref{app:Qk}, the
scale dependence of the halo bias in the quasi-linear regime is
described by a modified version of the empirical fitting function of
Tinker \etal (2005). This modification is needed to account for the
fact that we use a different definition of dark matter haloes, and is
characterized by one free `nuisance' parameter, $\psi$. In Paper~I we
calibrated $\psi$ using numerical $N$-body simulations of structure
formation in a $\Lambda$CDM cosmology, and found that $\psi = 0.9$ can
accurately fit the simulation results. In order to account for the
fact that we cannot rule out that $\psi$ is cosmology dependent, we
include an uncertainty of $\sim 15$ percent on $\psi$ in our
cosmological analysis. In particular, we treat $\psi$ as a free
parameter, but adopt a Gaussian prior centered on $\psi = 0.9$ and
with $\sigma_{\psi} = 0.15$.

For our fiducial model, we assume that the radial number density
distribution of satellite galaxies follows that of the dark matter
particles, i.e., $u_\rms(r|M) = u_\rmh(r|M)$. In other words, we
assume that satellite galaxies follow a NFW profile with the same
concentration-mass relation (Eq.~[\ref{concmass}]) as dark matter
haloes. In \S\ref{sec:nrsat} we relax this assumption and examine how
changes in $u_\rms(r|M)$ impact on our results.

Finally, we emphasize that the expression for the 1-halo term of the
galaxy-galaxy correlation function (Eq.~[\ref{P1h}]) has made the
implicit assumption that the halo occupation number of satellite
galaxies obeys Poisson statistics, i.e., that $\langle N_\rms (N_\rms
- 1) | M \rangle = \langle N_\rms |M \rangle^2$ (see Paper~I for
details). In \S\ref{sec:nonpoisson} we will relax this assumption, and
explore how deviations of $P(N_\rms|M)$ from a Poisson distribution
impact on our results.


\section{Methodology}
\label{sec:method}

The main goal of this paper is to obtain constraints on the
cosmological parameters $\Omega_\rmm$ and $\sigma_8$, and the halo
occupation distribution, as characterized by the CLF, using the SDSS
data described in \S\ref{sec:data} and the analytical model described
in \S\ref{sec:model}. We use Bayesian inference techniques to
determine the posterior probability distribution $P(\bolds\lambda |
\calD)$ of the model parameters $\bolds\lambda$, given the data
$\calD$.  According to Bayes' theorem,
\begin{equation}
P(\bolds\lambda | \calD)= {P(\calD | \bolds\lambda) \, P(\bolds\lambda) 
\over P(\calD)} \,,
\end{equation}
where $P(\calD | \bolds\lambda)$ is the {\it likelihood} of the data
given the model parameters, $P(\bolds\lambda)$ is the {\it prior
probability} of these parameters, and
\begin{equation}
P(\calD)= \int P(\calD | \bolds\lambda) \, P(\bolds\lambda) \, 
\rmd\bolds\lambda\,,
\end{equation}
is the {\it marginal probability} of the data, also called evidence
for the model. Since, we do not intend to perform model selection, the
evidence just acts as a normalization constant which need not be
calculated. Therefore the posterior distribution $P(\bolds\lambda|
\calD)$ is given by
\begin{equation}
P(\bolds\lambda | \calD) \propto \exp\left[
        \frac{-\chi^2(\bolds\lambda)}{2} \right]\,,
\end{equation}
where $\chi^2(\bolds\lambda)$ is a sum of the following terms
\begin{equation}\label{chisq}
\chi^2(\bolds\lambda) = \chi^2_{\rm LF} + \chi^2_{\rm Wp} +
  \chi^2_{\rm ESD} + \chi^2_{\rm prior}\,.    
\end{equation}
The first three terms quantify the goodness of the fit to the data and
correspond to the likelihood of the data given the parameters, while
the last term corresponds to the prior information we adopt. The
likelihood terms are given by
\begin{equation}
\chi^2_{\rm LF} = \sum_{j=1}^{32} \left[\frac{\Phi(L_j)-\tilde\Phi(L_j)}
{\sigma_\Phi(L_j)}\right]^2\,,
\end{equation}
\begin{equation}
\chi^2_{\rm Wp} = \sum_{k=1}^{6}
\left[\bw_\rmp(z_k)-\tilde\bw_\rmp(z_k)\right]^T \bC_{\rm Wp}^{-1}
\left[\bw_\rmp(z_k)-\tilde\bw_\rmp(z_k)\right]\,,
\end{equation}
\begin{equation}
\chi^2_{\rm ESD} = \sum_{k=1}^{6} \sum_{j=1}^{12}
\left[\frac{\Delta\Sigma(R_j,z_k) - \widetilde{\Delta\Sigma}(R_j,z_k)}
{\sigma_{\Delta\Sigma}(R_j,z_k)}\right]^2\,.
\end{equation}
Here $\tilde X$ denotes the model prediction for the observable $X$,
$\sigma_X$ is the corresponding error, $\bw_\rmp(z_k)$ is the vector of
the projected clustering measurements in the $k$--th luminosity bin,
and $\bC_{\rm Wp}$ is the covariance matrix of these measurements. 
\begin{figure}
\centerline{\psfig{figure=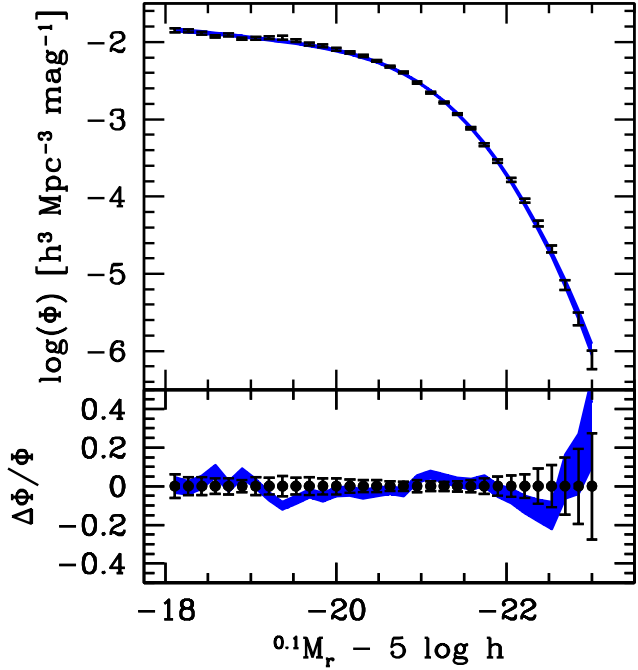,width=\hssize}}
\caption{{\it Upper panel:} The SDSS galaxy luminosity function in the
  $r$-band, $K$+$E$ corrected to $z=0.1$. Data points with error bars
  are taken from Blanton \etal (2003a), as described in
  \S\ref{sec:data}. The shaded region indicates the 95\% confidence
  levels of our \modF0 model, as obtained from the corresponding
  MCMC. {\it Lower panel:} The corresponding fractional errors,
  $\Delta \Phi/\Phi$.}
\label{fig:LF}
\end{figure}

As detailed in Paper I, our analytical model is accurate at the level
of 10 (in most cases 5) percent, in reproducing the 3-dimensional
galaxy-galaxy correlation and the galaxy-matter cross-correlation from
mock galaxy catalogs. Since the differences are not systematic, the
accuracy is expected to be much better for the projected galaxy-galaxy
correlation function, $w_\rmp(r_{\rm p})$, 
and the galaxy-galaxy lensing signal, $\Delta\Sigma(R)$. 
Therefore, we do not account for any systematic
uncertainty from our modeling in our likelihood estimate.

For our fiducial model, the set of model parameters, $\bolds\lambda$,
includes our primary cosmological parameters of interest,
$\Omega_\rmm$ and $\sigma_8$, the set of secondary cosmological
parameters $(\Ombh, n_\rms, h)$, the CLF parameters
$\bolds\lambda_{\rm CLF}$ and a set of nuisance parameters
$(\psi,\eta)$. Throughout we adopt uniform, non-informative priors on
our primary cosmological parameters as well as on all CLF parameters.
For the secondary cosmological parameters, $\bolds\lambda_{\rm sec} =
(\Ombh, n_\rms, h)$, we include priors from the WMAP7 analysis.  In
order to obtain the covariance matrix of these parameters ($\bC_{\rm
  sec}$), we have used the Monte Carlo Markov chains from the WMAP7
analysis (kindly provided to us by E. Komatsu) and marginalized over
all other parameters present in their analysis.  Finally, for the two
nuisance parameters we adopt Gaussian priors, as discussed in
\S\ref{sec:ingredients}. Hence, we have that
\begin{eqnarray}\label{prior}
\chi^2_{\rm prior} & = &
\left[\bolds\lambda_{\rm sec} - \tilde{\bolds\lambda}_{\rm sec}\right]^T 
\bC_{\rm sec}^{-1}
\left[\bolds\lambda_{\rm sec} - \tilde{\bolds\lambda}_{\rm sec}\right] +
\nonumber \\
& & \sum_i \left(\frac{\lambda_i-\tilde\lambda_i}{\sigma_{\lambda_i}}\right)^2 \,,
\end{eqnarray}
where the summation is over the two nuisance parameters $\psi$ and $\eta$.

We sample the posterior distribution of our model parameters given the
data using a Monte-Carlo Markov chain (MCMC). In particular, we
implement the Metropolis-Hastings algorithm to construct the MCMC
(Metropolis \etal 1953; Hastings 1970).  At any point in the chain, a
trial model is generated using a method specified below. The
chi-squared statistic for the trial model, $\chi^2_{\rm try}$, is
calculated using Equations (\ref{chisq})--(\ref{prior}).  This trial
model is accepted to be a member of the chain with a probability given
by
\begin{equation}
P_{\rm accept}=\left\{ \begin{array}{ll}
        1.0 & \mbox{if $\chi^2_{\rm try} \le \chi^2_{\rm cur}$} \\
  {\rm exp}[-(\chi^2_{\rm try}-\chi^2_{\rm cur})/2] & 
              \mbox{if $\chi^2_{\rm try} > \chi^2_{\rm cur}$}
\end{array}\right. \,,
\end{equation}
where $\chi^2_{\rm cur}$ denotes the $\chi^2$ for the current model in
the chain. We initialize the chain from a random position in our
multi-dimensional parameter space and obtain a chain of $\sim 50,000$
models. We discard the first $2000$ models (the {\it burn-in} period)
allowing the chain to sample from a more probable part of the
distribution. We use this chain of models to estimate the confidence
levels on the parameters and different observables of interest.

\begin{figure*}
\begin{center}$
\begin{array}{cc}
\psfig{figure=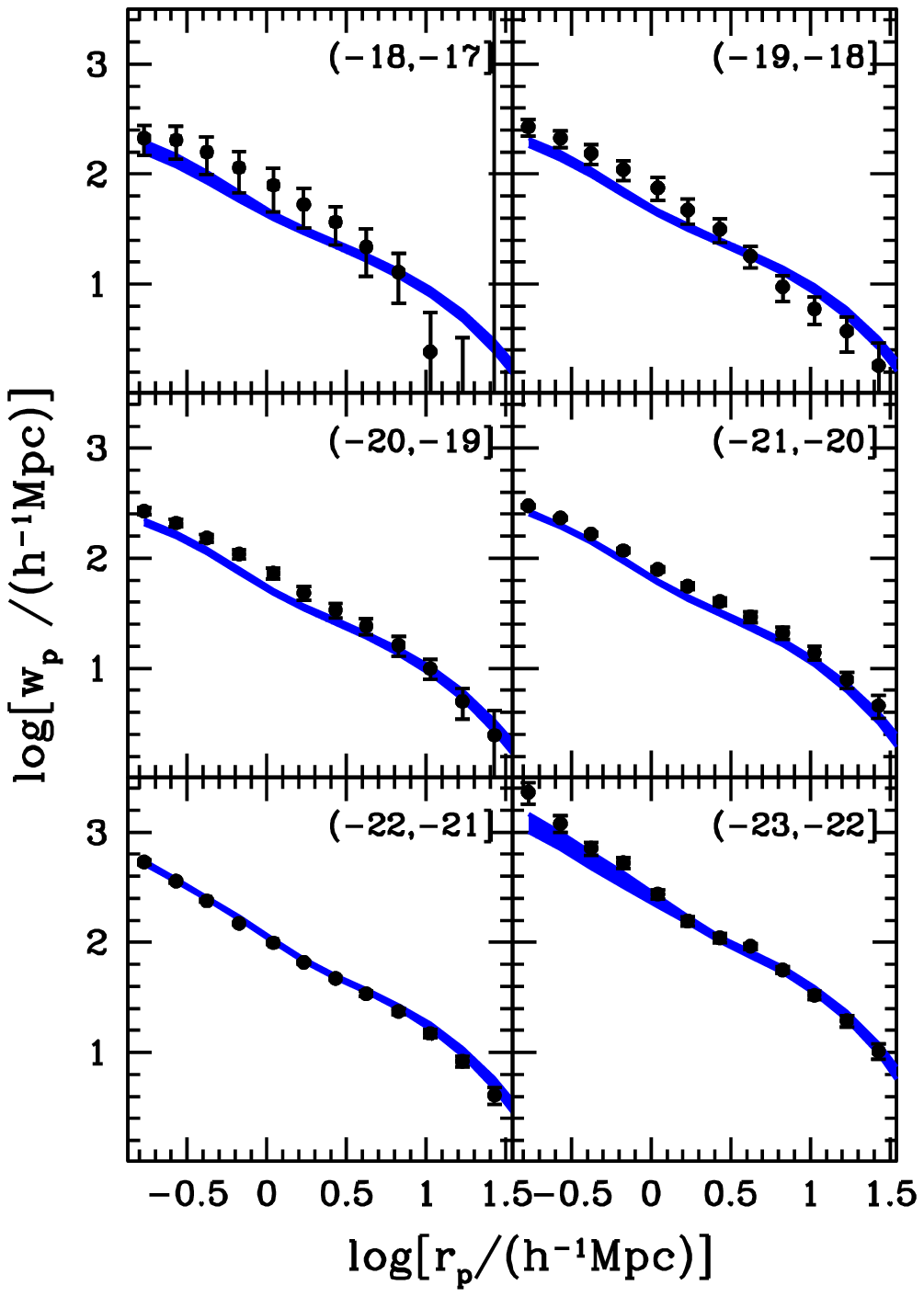,width=0.479\hdsize} &
\psfig{figure=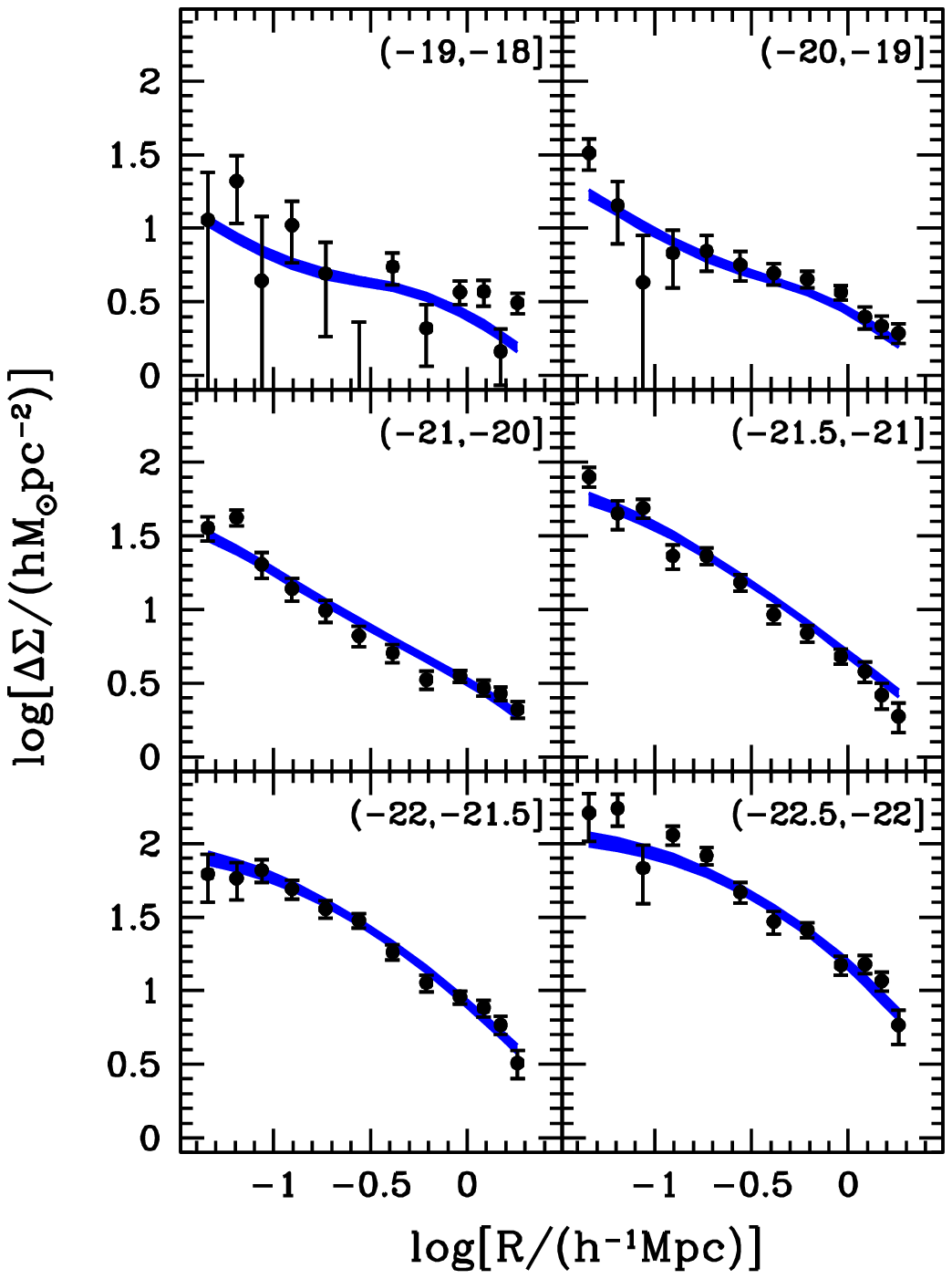,width=0.5\hdsize}  
\end{array}$
\end{center}
\caption{{\it Left-hand Panel.}  Projected correlation functions,
  $w_\rmp(r_\rmp)$, for the six different magnitude bins listed in
  Table~1, as indicated. Solid dots with ($1\,\sigma$) error bars are
  the SDSS data taken from Zehavi \etal (2011), while the shaded
  regions indicate the 95\% confidence levels obtained from the MCMC
  of our \modF0 model. We caution that, because of the covariance in the data, which is
accounted for in the modeling (see \S\ref{sec:method}), the quality of
the fit cannot be judged by eye. {\it Right-hand Panel.}  Excess surface density
  profiles, $\Delta\Sigma(R)$, for the six magnitude bins listed in
  Table~2, as indicated. Solid dots with ($1\,\sigma$) error bars are
  the SDSS data of Seljak \etal (2005) and Mandelbaum \etal (2006). As
  in the left-hand panels, the shaded regions indicate the 95\%
  confidence levels of our \modF0 model.}
\label{fig:ESDWp}
\end{figure*}

A proper choice of the proposal distribution is very important in
order to achieve fast convergence and a reasonable acceptance rate for
the trial models. The posterior distribution in a multi-dimensional
parameter space, such as the one we are dealing with, will have
numerous degeneracies and in general can be very difficult to sample
from. We have adopted the following strategy to overcome these
difficulties. During the first half of the burn-in stage, we chose an
independent Gaussian proposal distribution for every model parameter,
as is common for the Metropolis-Hastings algorithm.  Half-way through
the burn-in stage, we perform a Fisher information matrix analysis at
the best fit model found thus far. The Fisher information matrix,
given by
\begin{equation}
F_{ij} = -\frac{\partial^2 \ln\calL}{\partial \lambda_i \partial
    \lambda_j}\,,
\end{equation}
is a $N_\rmp \times N_\rmp$ symmetric matrix, where $N_\rmp$ denotes
the number of parameters in our model, and $\calL \propto e^{-\chi^2/2}$ is
the likelihood. The inverse of the Fisher matrix gives the covariance
matrix, $\bC_{\rm prop}$, of the posterior constraints on the model
parameters (see Paper~II)\footnote{The subscript `prop' indicates that
  this matrix is used to describe the proposal distribution.}. More
importantly, the eigenvectors of the covariance matrix are an
excellent guide to the numerous degeneracies in the posterior
distribution, and the corresponding eigenvalues set a scale for how
wide the posterior ought to be in a given direction. Therefore, for
the second half of the burn-in period, we utilize this information and
use a proposal distribution which is a multi-variate Gaussian centered
at the current value of the parameters and with a covariance equal to
$\bC_{\rm prop}$. In practice, the trial model ($\bolds\lambda_{\rm
  try}$) can be generated from the current model ($\bolds\lambda_{\rm
  cur}$) using
\begin{equation}
\bolds\lambda_{\rm try}=\bolds\lambda_{\rm cur} + \varsigma\,\bA \bx \,,
\end{equation}
where $\bx$ is a vector consisting of $N_\rmp$ standard normal
deviates, the matrix $\bA$ is such that $\bA\bA^T=\bC_{\rm prop}$, and
$\varsigma$ is a parameter that we have chosen to achieve an average
acceptance rate of $\sim 25$\%.  We repeat the Fisher matrix analysis
once again at the end of the burn-in period (using the best fit model
found thus far) and use the covariance matrix to define our proposal
distribution to be used for the MCMC. We have found this strategy to
be extremely efficient in sampling our posterior
distributions\footnote{A general-purpose python implementation of the
  MCMC sampler we have used in our work is available from the authors
  upon request.}.


\section{Results}
\label{sec:results}

Having described the data, the model, and the methodology, we now turn
to our results. In this section we describe the cosmological
constraints obtained for our \modF0 model, whereas
\S\ref{sec:systematics} discusses the robustness of these results to
model variations.  The constraints on galaxy bias, as characterized
via the CLF, are discussed in \S\ref{sec:CLFres}.
\begin{figure}
\centerline{\psfig{figure=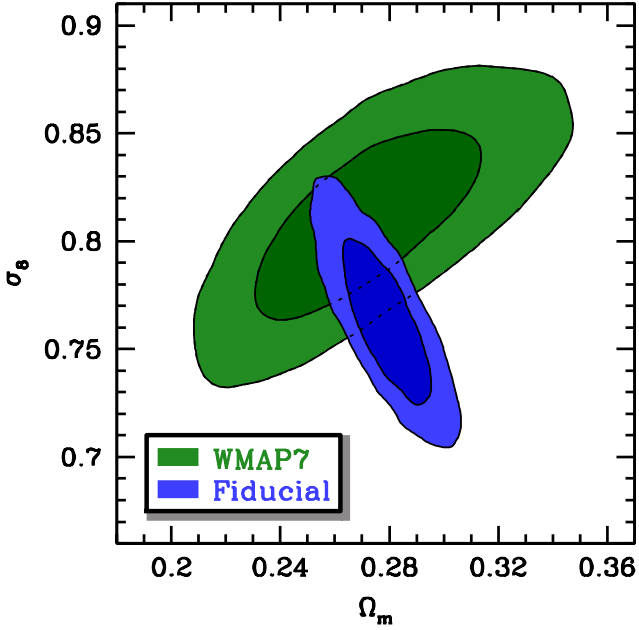,width=\hssize}}
\caption{The 68\% and 95\% CLs of the joint two-dimensional,
  marginalized posterior distribution for our \modF0 model, obtained
  from the simultaneous analysis of the abundance, clustering and
  lensing of galaxies in the SDSS. The green contours show the
  corresponding CLs for the WMAP7 analysis of the CMB (Komatsu \etal
  2011), and are shown for comparison.}
\label{fig:Fiducial}
\end{figure}

As discussed above, our \modF0 model consists of 16 free parameters;
the two primary cosmological parameters of interest, $\Omega_\rmm$ and
$\sigma_8$, for which we use uniform, non-informative priors, the
secondary cosmological parameters $n_\rms$, $h$ and $\Ombh$, for which
we use priors from WMAP7 (including their covariance), the 9 CLF
parameters that describe the halo occupation statistics, also with
uniform, non-informative priors, and finally the 2 nuisance
parameters, $\psi$ and $\eta$, for which we adopt Gaussian priors as
described in \S\ref{sec:ingredients}.  With a grand total of 182
constraints (32 data points for the LF, six bins of 13 data points
each for the projected correlation function\footnote{Although the
  galaxy-galaxy clustering data points have covariance, we have
  verified that the covariance matrix for each luminosity bin has rank
  equal to $13$, and therefore does not reduce the number of
  constraints.} and six bins of 12 ESD data points), this implies $182
- 16 + 5 = 171$ degrees of freedom, which is the number we have used
to compute the reduced $\chi^2$ values listed in the final column of
Table~3.

Figs.~\ref{fig:LF} and~\ref{fig:ESDWp} compare the predictions of the
\modF0 model (shaded regions, indicating the 95\% confidence levels)
to the data used to constrain the model (solid dots with error
bars, indicating the 68 \% confidence levels). 
Fig.~\ref{fig:LF} shows that the model accurately fits the
$r$-band galaxy luminosity function.  Although most data points agree
with the model predictions at the $1\,\sigma$ level, the data reveals
a few small `wiggles' at the faint end that are not reproduced by the
model, and which contribute dominantly to $\chi^2_{\rm LF}$, the value
of which is listed in Table~3.
\begin{figure*}
\centerline{\psfig{figure=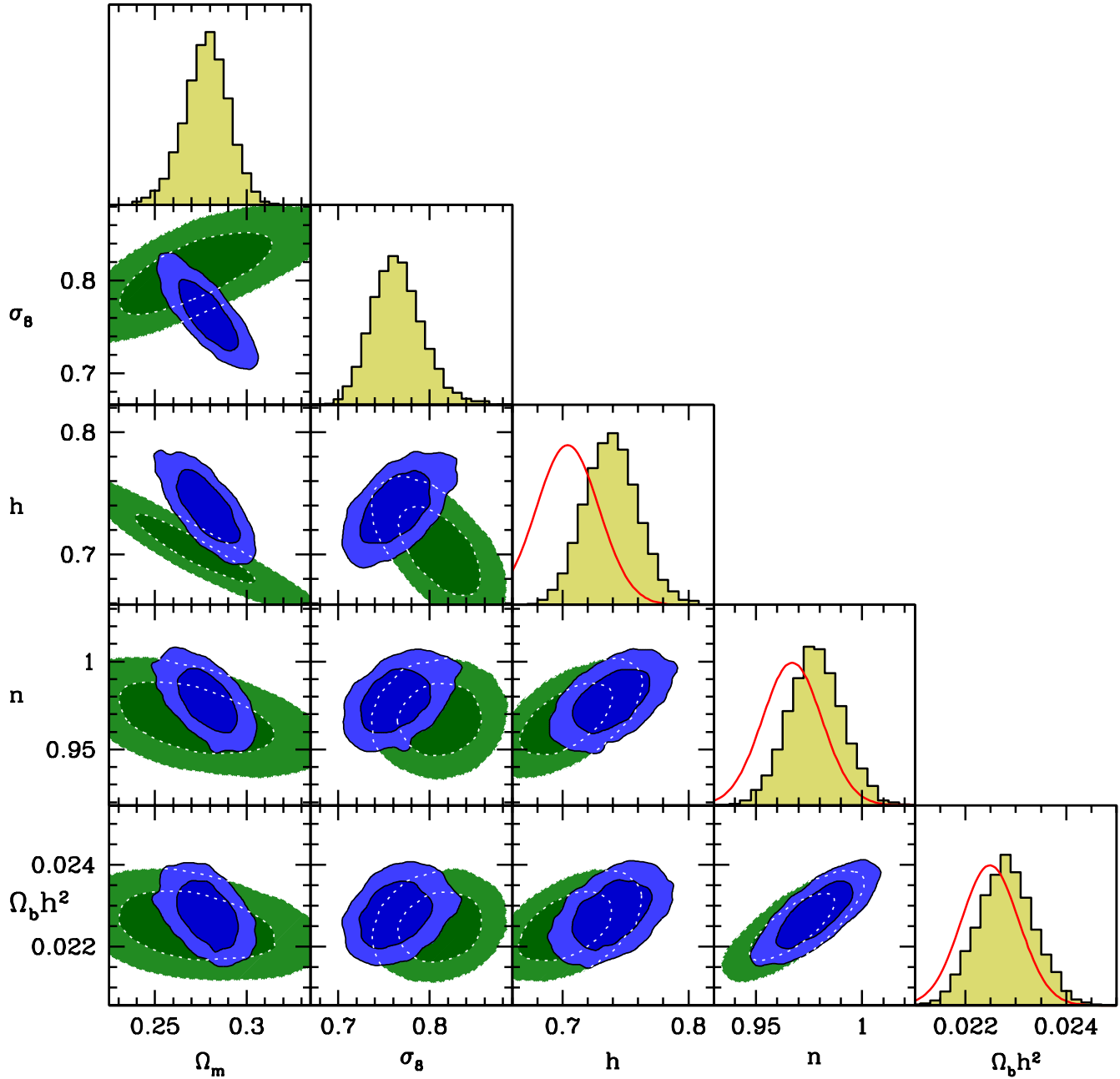,width=\hdsize}}
\caption{Cosmological constraints for the \modF0 model.  Histograms
  show the marginalized posterior distributions, while the blue
  contour show the 68\% and 95\% CLs of the joint, two-dimensional marginalized
  posterior distributions. For comparison, the corresponding CLs from
  WMAP7 are shown as green contours, while the red, solid curves show
  the marginalized WMAP7 prior distributions used for the secondary
  cosmological parameters, $n_\rms$, $h$ and $\Ombh$.}
\label{fig:CosmoHisto}
\end{figure*}

The left-hand side of Fig.~\ref{fig:ESDWp} shows the projected
correlation functions, $w_\rmp(r_\rmp)$, for six different magnitude
bins. We caution that, because of the covariance in the data, which is
accounted for in the modeling (see \S\ref{sec:method}), the quality of
the fit cannot be judged by eye. However, it is evident from the
$\chi^2$ values of the best-fit \modF0 model (see Table~3), that the
total $\chi^2$ is clearly dominated by $\chi^2_{\rm Wp}$. In
particular, $\chi^2_{\rm Wp} > 2\chi^2_{\rm ESD}$, even though the
projected correlation functions only have $78/72 \simeq 1.08$ times as
many data points.  It turns out $\chi^2_{\rm Wp}$ is dominated by the
contribution from the data in the $[-20,-21]$ magnitude
bin. Interestingly, this bin covers the volume that encloses the Sloan
Great Wall (SGW), a huge supercluster at $z \sim 0.08$ and the largest
coherent structure detected in the SDSS (Gott \etal 2005). As
discussed in Zehavi \etal (2011), pruning the data sample so as to
exclude the SGW region results in a significantly reduced clustering
strength for galaxies in the $[-20,-21]$ magnitude range (i.e., the
correlation length is reduced from $r_0 = 5.46 \pm 0.15$ to $4.82 \pm
0.23$). We return to this issue, and its potential impact on our
cosmological constraints, in \S\ref{sec:variance} when we discuss the
potential impact of sample variance.

Finally, the right-hand side of Fig.~\ref{fig:ESDWp} shows the excess
surface densities, $\Delta\Sigma(R)$, again for six different
magnitude bins as indicated. The model nicely reproduces the overall
trends in the data, with only a few data points that fall outside the
95\% confidence region of the model. Overall, we conclude that our
\modF0 model is consistent with the data at a satisfactory level.
In particular, the most important features in the data are nicely
reproduced by the model and find a natural explanation within the
framework of the halo model. For example, the fact that brighter
galaxies reveal stronger clustering and higher excess surface densities
is consistent with the common notion that brighter galaxies reside in
more massive haloes. The lensing signal is directly sensitive to this
aspect because it probes the matter distribution around galaxies,
whereas the clustering signal is affected by it only indirectly due to
the fact that more massive haloes are more strongly clustered than
less massive ones (e.g., Mo \& White 1996). Also, the relatively weak
deviations of $w_\rmp(r_\rmp)$ and $\Delta\Sigma(R)$ from pure
power-laws typically reflect transitions from scales where the signal
is dominated by different components of the power spectra. Examples
are the 1-halo to 2-halo transition (e.g., Zehavi \etal 2004) and the
1-halo central to 1-halo satellite transition for the excess surface
densities (e.g., Cacciato \etal 2009).

\subsection{Cosmological Parameters}
\label{sec:FidCosmo}

Fig.~\ref{fig:Fiducial} shows the constraints on our two primary
cosmological parameters of interest; $\Omega_\rmm$ and $\sigma_8$. The
blue contours show the 68\% and 95\% CLs of the joint two-dimensional,
marginalized posterior distribution obtained from our simultaneous
analysis of the abundance, clustering and lensing of galaxies in the
SDSS. The green contours show the corresponding CLs for the WMAP7
analysis of the CMB (Komatsu \etal 2011), and are shown for
comparison. Note that our results are in excellent agreement with
those from WMAP7, strengthening the case for a true concordance
cosmology. In particular, our analysis yields $\Omega_\rmm =
0.278_{-0.026}^{+0.023}$ and $\sigma_8 = 0.763_{-0.049}^{+0.064}$
(both 95\% CL), while the WMAP7 analysis has $\Omega_\rmm =
0.264_{-0.049}^{+0.064}$ and $\sigma_8 = 0.801_{-0.058}^{+0.059}$
(both 95\% CL). Note also that the degeneracy between $\Omega_\rmm$
and $\sigma_8$ inherent in our analysis runs perpendicular to that
inherent in the CMB data. This indicates that a combined analysis will
be able to significantly tighten the constraints on $\Omega_\rmm$ and
$\sigma_8$ (see also Paper~II).  Finally, Fig.~\ref{fig:Fiducial}
suggests that our constraints are even tighter than those from the
WMAP7 analysis. However, we emphasize that this is not a fair
comparison since we have used priors from WMAP7 on the secondary
cosmological parameters $n_\rms$, $h$ and $\Ombh$, but not on
$\Omega_\rmm$ or $\sigma_8$ (see Paper~II for the case with no priors 
on $n_\rms$, $h$ and $\Ombh$). 

Fig.~\ref{fig:CosmoHisto} shows the one-dimensional (histograms) and
joint two-dimensional (contour plots) marginalized posterior
distributions on all five cosmological parameters.  Solid contours
indicate the 68\% and 95\% CLs obtained from the analysis presented
here, while the dotted contours are the 68\% and 95\% CLs from the
WMAP7 analysis, shown for comparison.  The strongest parameter
degeneracies are between $\Omega_\rmm$ and $\sigma_8$
(cross-correlation coefficient $r=-0.81$), between $\Ombh$ and
$n_\rms$ ($r=0.79$), and between $\Omega_\rmm$ and $h$
($r=-0.74$). All other combinations are only weakly correlated with
$\vert r \vert < 0.5$.

Overall, there is good agreement between our constraints and those
inferred from the WMAP7 data (see also Table~4).  However, there is
some tension regarding the secondary cosmological parameters, which is
evident from the fact that the posterior and prior distributions
(indicated by red, solid curves) reveal an offset. This is most
pronounced for the Hubble parameter $h$; whereas the WMAP7 prior used
has $h = 0.710^{+0.050}_{-0.049}$ our posterior distribution has $h =
0.739^{+0.044}_{-0.041}$ (both 95\% CL).  For comparison, the revised
parallaxes for Cepheid stars by van Leeuwen \etal (2007) raises the
value for $h$ from the HST Key Project from $0.72 \pm 0.08 \kmsmpc$
(Freedman \etal 2001) to $0.76 \pm 0.08 \kmsmpc$ (68\% CL), and the
Cepheid-based determination of Sandage \etal (2006) from $0.62 \pm
0.05 \kmsmpc$ to $0.70 \pm 0.05 \kmsmpc$ (68\% CL). Hence, despite
some tension with the WMAP7 based constraints, our posterior
distribution for the Hubble constant is well within the range of
values inferred from Cepheids. Interestingly, our constraint on $h$ is
correlated with the constraints on both $\Omega_\rmm$ and $\sigma_8$:
according to our analysis, a relatively low Hubble parameter of $h
\sim 0.7$ implies $\Omega_\rmm \sim 0.30$ and $\sigma_8 \sim 0.73$,
while for $h\sim 0.78$ the results presented here suggest that
$\Omega_\rmm \sim 0.25$ and $\sigma_8 \sim 0.82$. Clearly, an improved
constraint for the Hubble parameter could help to significantly
tighten the constraints on both $\Omega_\rmm$ and $\sigma_8$.

\subsection{Nuisance parameters}
\label{sec:FidNuisance}

As discussed in \S\ref{sec:ingredients}, our model contains two
nuisance parameters: $\psi$, which enters in the description of the
scale dependence of the halo bias, and $\eta$, which sets the
normalization of the halo concentration-mass relation. As discussed in
detail in Paper~I, the freedom in $\eta$ also characterizes model
uncertainties arising from our oversimplifications regarding the
structure of dark matter haloes.

\begin{figure}
\centerline{\psfig{figure=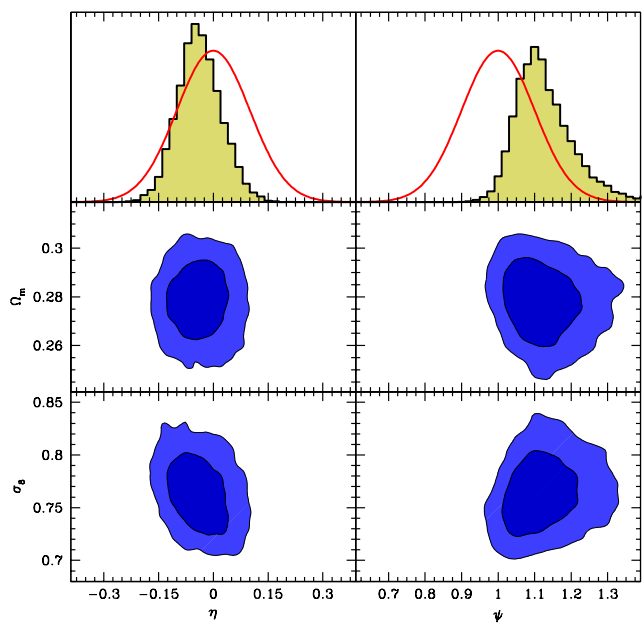,width=\hssize}}
\caption{The shaded histograms in the upper panels show the
  marginalized posterior distributions for the two nuisance parameters
  $\eta$ (left-hand panel), which sets the normalization of the halo
  concentration-mass relation, and $\psi$ (right-hand panel), which
  enters in the description of the scale dependence of the halo bias.
  The solid, red curves show the corresponding Gaussian prior
  distributions. The contour plots in the middle and lower panels show
  the joint, two-dimensional marginalized posterior distributions of
  $\Omega_\rmm$ and $\eta$ (middle left-hand panel), $\sigma_8$ and
  $\eta$ (lower left-hand panel), $\Omega_\rmm$ and $\psi$ (middle
  right-hand panel) and $\sigma_8$ and $\psi$ (lower right-hand
  panel). See \S\ref{sec:FidNuisance} for a detailed discussion.}
\label{fig:nuisance}
\end{figure}

The upper panels of Fig.~\ref{fig:nuisance} show the posterior
distributions for $\eta$ (left-hand) and $\psi$ (right-hand) of our
\modF0 model. The contours in the other panels show 68\% and 95\%
confidence levels of the joint two-dimensional marginalized posteriors
with $\Omega_\rmm$ and $\sigma_8$.  Clearly, neither $\eta$ nor $\psi$
shows significant correlation with $\Omega_\rmm$ or $\sigma_8$ (in all
cases the cross-correlation coefficient $\vert r \vert < 0.3$).  As
already discussed in Paper~II, this is an important result, as it
indicates that the uncertainties in the scale dependence of the halo
bias and the oversimplifications regarding the structure of dark
matter haloes do not have a significant impact on the cosmological
constraints presented here. 

The solid, red curves in the upper panels of Fig.~\ref{fig:nuisance}
reflect the Gaussian priors that we imposed on our model. In the case
of $\eta$, the posterior distribution ($\eta = -0.04^{+0.13}_{-0.11}$,
95\% CL) is significantly narrower than the prior distribution ($\eta
= 0.0^{+0.2}_{-0.2}$, 95\% CL), indicating that the prior did not have
a significant impact on our results. In the case of $\psi$, however,
the posterior distribution ($\psi = 1.12^{+0.20}_{-0.12}$, 95\% CL) is
clearly offset from the prior ($\psi = 0.90^{+0.30}_{-0.30}$, 95\% CL)
to larger values. This might raise concern that a less restrictive
prior might have resulted in significantly different cosmological
constraints.  However, this is not the case for the following two
reasons. First, as already mentioned above, and discussed in more
detailed in Paper~II, $\psi$ is only very poorly correlated with the
cosmological parameters. Second, as discussed in the Appendix, when
$\psi$ increases the radial bias function $\zeta(r,z)$ asymptotes to
the empirical fitting function of Tinker \etal (2005). In other words,
once $\psi$ increases beyond a certain value, any further increase has
zero impact. For the best-fit cosmology of our \modF0 model, this
critical value of $\psi$ is $\sim 1.5$.

\subsection{The Galaxy-Dark Matter Connection}
\label{sec:CLFres}

\begin{figure*}
\psfig{figure=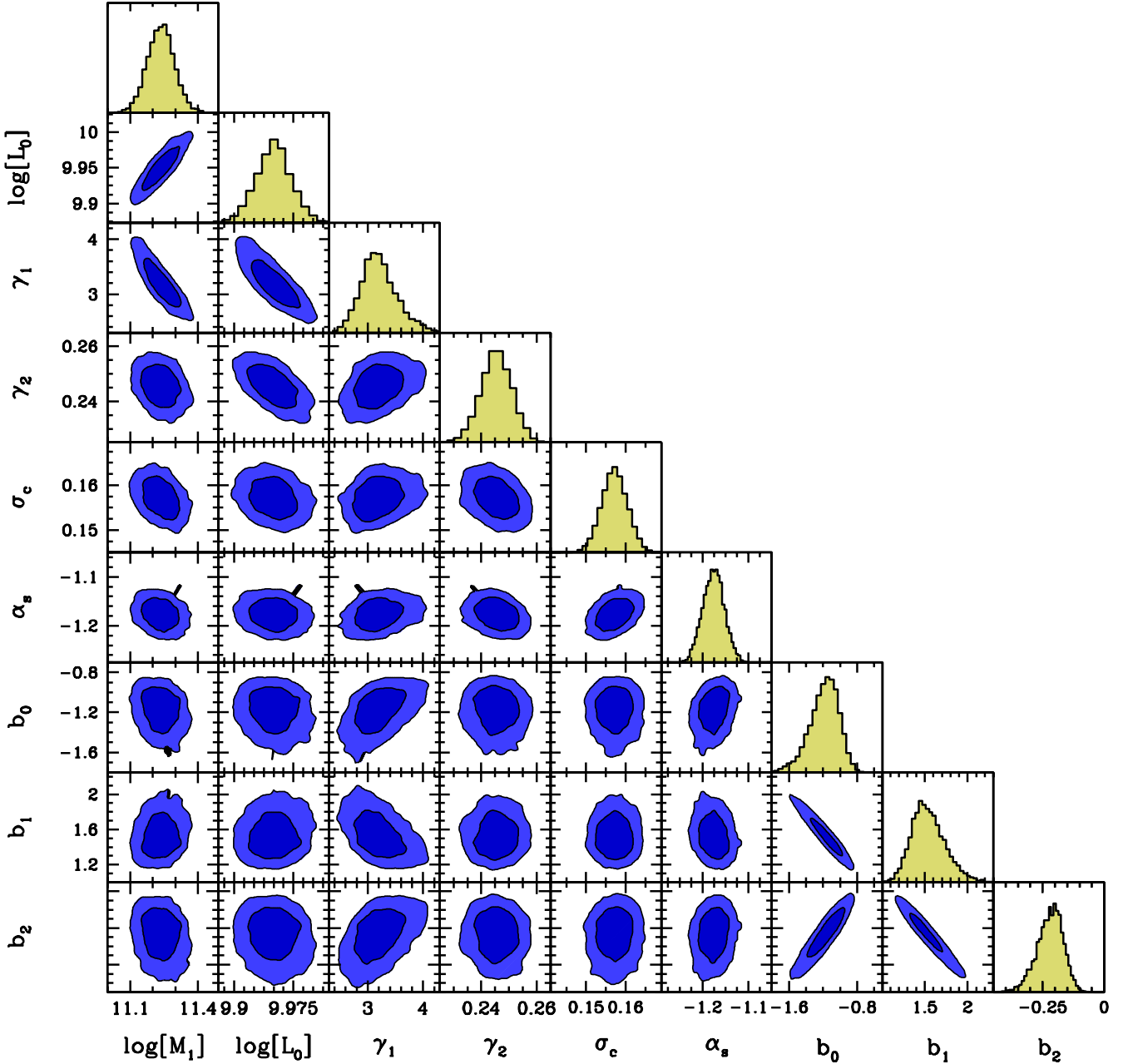,width=\hdsize}
\caption{Constraints on the CLF parameters for the \modF0 model.
  Histograms show the marginalized posterior distributions, while the
  blue contour show the 68\% and 95\% CLs of the joint, two-dimaensional 
  marginalized posterior distributions.}
\label{fig:CLFparam}
\end{figure*}

One of the powerful aspects of the method used here, is that the data
is used to simultaneusly constrain cosmology and halo occupation
statistics.  Fig.~\ref{fig:CLFparam} shows the one-dimensional
(histograms) and joint two-dimensional (contour plots) marginalized
posterior distributions of model \modF0 for the nine CLF parameters
that describe the relation between galaxy luminosity and halo
mass. The medians and 95\% CLs are also listed in Table~5.  All nine
parameters are tightly constrained, with tight degeneracies between
the parameters $b_0$, $b_1$ and $b_2$, that describe the normalization
of the satellite CLF (see Eq.~[\ref{phi}]), and between the parameters
$L_0$, $M_1$ and $\gamma_1$, that describe the relation between halo
mass and the luminosity of its central galaxy (see Eq.~[\ref{LcM}]).

The constraint on the faint-end slope of the satellite CLF is
$\alpha_\rms = -1.18 \pm 0.05$ (95\% CL), in good agreement with
results obtained from galaxy group catalogues (e.g., Eke \etal 2004;
Yang \etal 2008). The constraints on the scatter in the CLF of central
galaxies is $\sigma_\rmc = 0.157 \pm 0.007$ (95\% CL), which is in
excellent agreement with a variety of other constraints, from
satellite kinematics (More \etal 2009), from clustering and lensing
(Cacciato \etal 2009; Moster \etal 2010), and from galaxy group
catalogues (Yang \etal 2008). Interestingly, such an amount of scatter
is also in excellent agreement with predictions from semi-analytical
models for galaxy formation (e.g., Wang \etal 2006; see also More
\etal 2009).

\begin{figure*}
\begin{center}
\psfig{figure=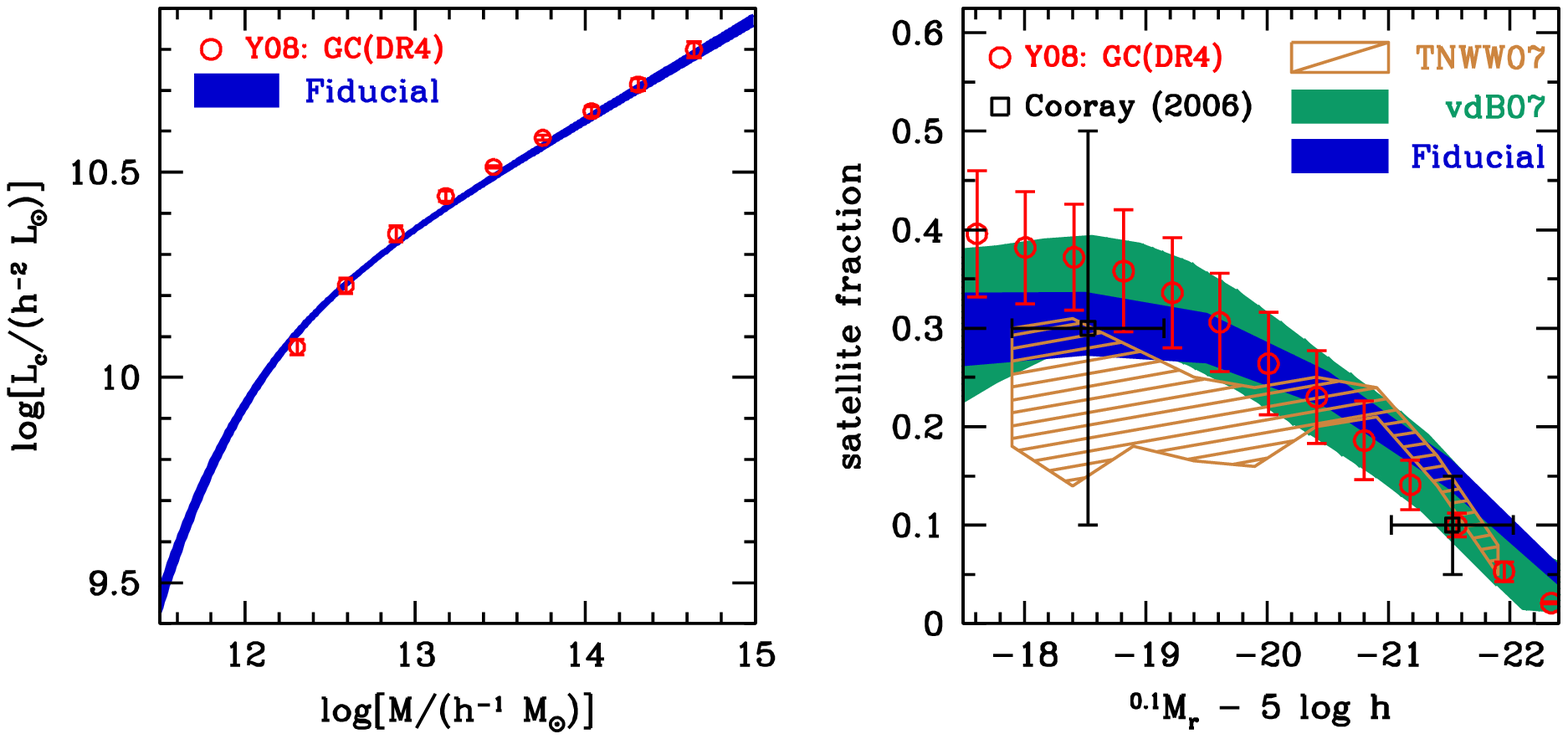,width=\hdsize}
\end{center}
\caption{ {\it Left-hand panel.}  The luminosity of central galaxies,
  $L_\rmc$, as function of halo mass, $M$. The shaded region indicates
  the 68\% confidence interval for the \modF0 model, while the 
  open red circles indicate the results of Yang \etal (2008) obtained from
  the SDSS group catalogue of Yang \etal (2007; error bars indicate the 68 \% CL),
  where we properly converted their group masses to our definition of halo mass and the
  cosmology of the best-fit \modF0 model. {\it Right-hand panel.}  The
  satellite fractions as a function of $r$-band magnitude. The blue
  shaded regions indicate the 68 \% CL for the \modF0
  model. This is compared to the satellite fractions obtained by
  Cooray (2006; black, open squares) from a CLF analysis of the SDSS,
  by Yang \etal (2008; open red circles) from the SDSS group catalogue
  (error bars indicate the 68 \% CL in both cases), by van den Bosch \etal
  (2007; green shaded area, indicating the 68\% CL) and
  Tinker \etal (2007; orange shaded area, indicating the 68\% CL), 
  both from halo occupation modeling of galaxies
  in the 2dFGRS.}
\label{fig:CLFcomp}
\end{figure*}

In the left-hand panel of Fig.~\ref{fig:CLFcomp} we compare the
constraints on the $L_\rmc$-$M$ relation of our \modF0 model
(shaded area, indicating the 68\% confidence region) to the results
obtained by Yang \etal (2008) from the SDSS galaxy group catalogue
(symbols with errorbars indicating the 68\% confidence level).  Here
we have converted the group masses listed in their Table~1 to our
definition of halo mass and the cosmology of the best-fit \modF0 model
using the abundance matching technique described in Eq.~(15) of Yang
\etal (2007).  Finally, the right-hand panel of Fig.~\ref{fig:CLFcomp}
compares the satellite fractions as function of galaxy luminosity
obtained from our \modF0 model (blue shaded area, indicating the 68\%
CL) to constraints obtained by Cooray (2006; black, open
squares), Yang \etal (2008; red, open circles), van den Bosch \etal
(2007; green shaded area, indicating the 68\% CLs), and
Tinker \etal (2007; orange shaded area, indicating the 68\% CL). 
All these constraints are obtained comparing halo occupation models to data from the SDSS or
2dFGRS, and are in excellent agreement with each other and with the
constraints from our \modF0 model.

Based on all these comparisons, we conclude that the constraints on
the galaxy-dark matter connection for our \modF0 model are in
excellent agreement with a wide variety of independent constraints.
We emphasize that, contrary to many other studies, 
our combined  analysis of abundance, clustering, 
and lensing of galaxies also accounts for uncertainties in cosmological parameters. 
This strongly supports that our method of simultaneously constraining
cosmology and halo occupation statistics yields reliable results.
\begin{table*}\label{tab:cosmores}
\caption{Constraints on Cosmological Parameters}
\begin{tabular}{cccccc}
\hline\hline
Model & $\Omega_\rmm$ & $\sigma_8$ & $h$ & $n_\rms$ & $100\,\Ombh$ \\
  (1) & (2) & (3) & (4) & (5) & (6) \\
\hline
\modF0  & $0.278_{-0.026}^{+0.023}$ & $0.763_{-0.049}^{+0.064}$ & $0.739_{-0.041}^{+0.044}$ & $0.978_{-0.025}^{+0.026}$ & $2.279_{-0.107}^{+0.112}$ \\
\modG0  & $0.279_{-0.026}^{+0.025}$ & $0.761_{-0.049}^{+0.061}$ & $0.738_{-0.043}^{+0.042}$ & $0.977_{-0.026}^{+0.025}$ & $2.279_{-0.110}^{+0.103}$ \\
\modD10 & $0.280_{-0.023}^{+0.023}$ & $0.744_{-0.047}^{+0.056}$ & $0.741_{-0.040}^{+0.040}$ & $0.980_{-0.026}^{+0.026}$ & $2.287_{-0.109}^{+0.111}$ \\
\modJ0b & $0.263_{-0.031}^{+0.026}$ & $0.805_{-0.061}^{+0.083}$ & $0.727_{-0.042}^{+0.054}$ & $0.977_{-0.027}^{+0.030}$ & $2.278_{-0.112}^{+0.117}$ \\
\modK0  & $0.282_{-0.026}^{+0.026}$ & $0.727_{-0.049}^{+0.054}$ & $0.739_{-0.040}^{+0.042}$ & $0.979_{-0.023}^{+0.025}$ & $2.286_{-0.101}^{+0.102}$ \\
\hline
WMAP7   & $0.264_{-0.049}^{+0.064}$ & $0.801_{-0.058}^{+0.059}$ & $0.710_{-0.049}^{+0.050}$ & $0.963_{-0.028}^{+0.029}$ & $2.257_{-0.113}^{+0.120}$ \\
\hline\hline
\end{tabular}
\medskip
\begin{minipage}{\hdsize}
  The cosmological constraints for the various models discussed in the
  text (see also Table~3).  Values indicate the median and 95\% CLs of
  the marginalized posterior distributions obtained from our
  MCMCs. For comparison, the final row lists the constraints obtained
  from the WMAP7 analysis of the CMB (taken from Komatsu \etal 2011).
\end{minipage}
\end{table*}
%


\section{Systematics}
\label{sec:systematics}

The \modF0 model, discussed in the previous section, relies on two
assumptions regarding satellite galaxies that may not be entirely
accurate. These concern the radial number density distribution of
satellites and the Poisson nature of the satellite occupation
numbers. In this section we gauge the impact of these assumptions on
our results. In addition, we also address the importance of properly
accounting for the residual redshift space distortions present in the
projected correlation functions used to constrain the models.  We do
so by running a series of MCMCs in which we include small variations
with respect to our \modF0 model. These models, and their respective
$\chi^2$-values for the best-fit model in the chain, are listed in
Table~3, while the cosmological constraints are summarized in
Table~4.

\subsection{Radial Number Density Profile of Satellites}
\label{sec:nrsat}

In our \modF0 model we have assumed that the radial number density
distribution of satellite galaxies in a halo of mass $M$ follows an
NFW profile with a concentration parameter that is identical to that
of the dark matter density profile, i.e., we assumed that $u_\rms(r|M)
= u_\rmh(r|M)$ (see \S\ref{sec:powerspectra}). This has observational
support from a number of studies of the radial distribution of
satellite galaxies in groups and clusters (e.g., Beers \& Tonry 1986;
Carlberg, Yee \& Ellingson 1997a; van der Marel \etal 2000; Biviano \&
Girardi 2003; Lin, Mohr \& Stanford 2004; van den Bosch \etal 2005b;
Sheldon \etal 2009).

However, a number of recent studies have argued that the spatial
distribution of satellite galaxies is less centrally concentrated than
the dark matter (e.g., Yang \etal 2005; Chen 2008; More \etal 2009).
Interestingly, an opposite result has been obtained for luminous red
galaxies (e.g., Masjedi \etal 2006; Watson \etal 2010, 2012; Tal, Wake
\& van Dokkum 2012).  In fact, there are indications that the radial
distribution of satellite galaxies appears to have some dependence on
the color and/or morphologies of the central (e.g., Lorrimer \etal
1994; Sales \& Lambas 2005; Lares, Lambas \& Dominguez 2011; Guo \etal
2012).  From a theoretical point of view, one expects the radial
distribution of satellite galaxies to reflect that of dark matter
subhaloes. Numerical simulations have shown that subhaloes do not
follow the same spatial distribution as the dark matter; subhaloes
tend to populate preferentially the outskirts of their host haloes
(e.g., Diemand, Moore \& Stadel 2004; Gao \etal 2004; Springel \etal
2008). However, the stellar components of satellite galaxies may be
more resilient against tidal disruption than their dark matter
components, such that the radial profile of satellite galaxies is not
necessarily well represented by that of dark matter subhaloes in pure
$N$-body simulations.

In conclusion, the simple assumption that $u_\rms(r|M) = u_\rmh(r|M)$
is likely oversimplified. To address the potential impact of this
assumption on our results, we now consider a model in which we adopt
\begin{equation}\label{defcalR}
u_\rms(r|M) \propto \left({r \over \calR r_*}\right)^{-1}
\left(1+{r\over\calR  r_*}\right)^{-2} \,.
\end{equation}
Here $r_*$ is the scale radius of the dark matter density profile for
a halo of mass $M$, and $\calR$ is a free parameter.  For $\calR = 1$
this profile reduces to that of our \modF0 model, i.e., $u_\rms(r|M)
= u_\rmh(r|M)$. Hence, $\calR$ is a parameter that controls how
satellites are concentrated with respect to the dark matter. In order
to gauge how our results depend on $\calR$, and whether the data
prefers values of $\calR$ that differ from unity, we run a MCMC in
which $\calR$ is a free parameter for which we adopt Gaussian prior
with mean equal to unity and standard deviation, $\sigma_\calR =
0.25$. In what follows we refer to this MCMC as our \modG0 model.

As is apparent from the $\chi^2$ values of the best-fit model in the
\modG0 chain (see Table~3), the extra freedom in the model does not
result in a significantly better fit to the data.  More importantly,
the cosmological constraints are also unaffected (see upper left-hand
panel of Fig.~\ref{fig:alto} and Table~4). This is also evident from
the contour plots in the left-hand column of Fig.~\ref{fig:CosmoGal},
which show the 68 and 95\% confidence intervals of the joint
two-dimensional posterior distributions for $\calR$ and $\Omega_\rmm$
(middle row) and for $\calR$ and $\sigma_8$ (bottom row).  With
cross-correlation coefficients $\vert r \vert < 0.1$, it is clear that
neither $\Omega_\rmm$ nor $\sigma_8$ is significantly influenced by
potential uncertainty in $\calR$. Finally, the histogram in the upper
left-hand panel of Fig.~\ref{fig:CosmoGal} shows the posterior
distribution for $\calR$.  With $\calR = 0.96_{-0.29}^{+0.32}$ (95\%
CL) it is clear that the posterior distribution is narrower than the
prior distribution (solid, red curve), and that the data prefers
values of $\calR$ that are close to our fiducial value of unity. We
conclude that our results are robust to uncertainties in the radial
number density profile of satellite galaxies, and that the data used
here is consistent with satellite galaxies being an unbiased tracer of
the mass distribution within their host haloes (i.e., $\calR = 1$).
Interestingly, as mentioned before, this result is supported by studies
of satellite galaxies in groups and clusters 
(e.g., Beers \& Tonry 1986;
Carlberg, Yee \& Ellingson 1997a; van der Marel \etal 2000; Biviano \&
Girardi 2003; Lin, Mohr \& Stanford 2004; van den Bosch \etal 2005b;
Sheldon \etal 2009)
but it is somewhat in disagreement with a number of studies which indicate
that the radial distribution of satellite is less centrally concentrated than dark matter
(e.g., Yang \etal 2005; Chen 2008; More \etal 2009). 
It is worth mentioning that results based on groups might be potentially affected by 
the offset of central galaxies (see e.g. Skibba et al. 2011).
Clearly, more dedicated studies are required to further constrain the radial 
distribution of satellite galaxies.
 
\subsection{Poisson Statistics of Satellite Galaxies}
\label{sec:nonpoisson}

In our \modF0 model we have assumed that the occupation
numbers of satellite galaxies obey Poisson statistics, i.e.,
\begin{equation}
P(N_\rms|M) = {\langle N_\rms|M\rangle^{N_\rms} \, 
e^{-\langle N_\rms|M\rangle} \over N_\rms !}\,,
\end{equation}
such that
\begin{equation}\label{defAp}
\calA_\rmP \equiv {\langle N_\rms (N_\rms-1)|M\rangle \over 
\langle N_\rms|M \rangle^2} = 1\,.
\end{equation}
As shown in Yang \etal (2008), this assumption has strong support from
galaxy group catalogs. Additional support comes from numerical
simulations which show that dark matter subhaloes (which are believed
to host satellite galaxies) also follow Poisson statistics (e.g.,
Kravtsov \etal 2004). However, recently there have been several claims
that the occupation statistics of subhaloes and/or satellite galaxies
may deviate slightly, but significantly, from Poisson, i.e.,
$\calA_\rmP \ne 1$ (e.g., Porciani, Magliocchetti \& Norberg 2004; van
den Bosch \etal 2005a; Giocoli \etal 2010; Boylan-Kolchin \etal 2010;
Busha \etal 2011). As shown in Paper~I, the satellite-satellite term
of the 1-halo power spectrum scales linearly with $\calA_\rmP$, i.e.,
$P^{\rm 1h}_{\rm gg}(k) \propto \calA_\rmP$. Hence, any deviation of
$\calA_\rmP$ from unity has a direct impact on the projected
correlation functions $w_\rmp(r_\rmp)$ on small scales, at least if
the satellite-satellite term dominates over the central-satellite term
(which is independent of $\calA_\rmP$).

In order to quantify how our results depend on uncertainties related
to the exact form of $P(N_\rms|M)$, we run a MCMC in which we include
$\calA_\rmP$ as a free parameter. Motivated by the empirical results
of Yang \etal (2008), which suggest that the occupation statistics of
satellite galaxies in the SDSS are not too different from Poisson,
we adopt a Gaussian prior for $\calA_\rmP$ with mean equal to unity
and standard deviation, $\sigma_{\calA_\rmP} = 0.1$. In what follows
we refer to this MCMC as our \modD10 model.

\begin{figure*}
\begin{center}$
\begin{array}{cc}
\psfig{figure=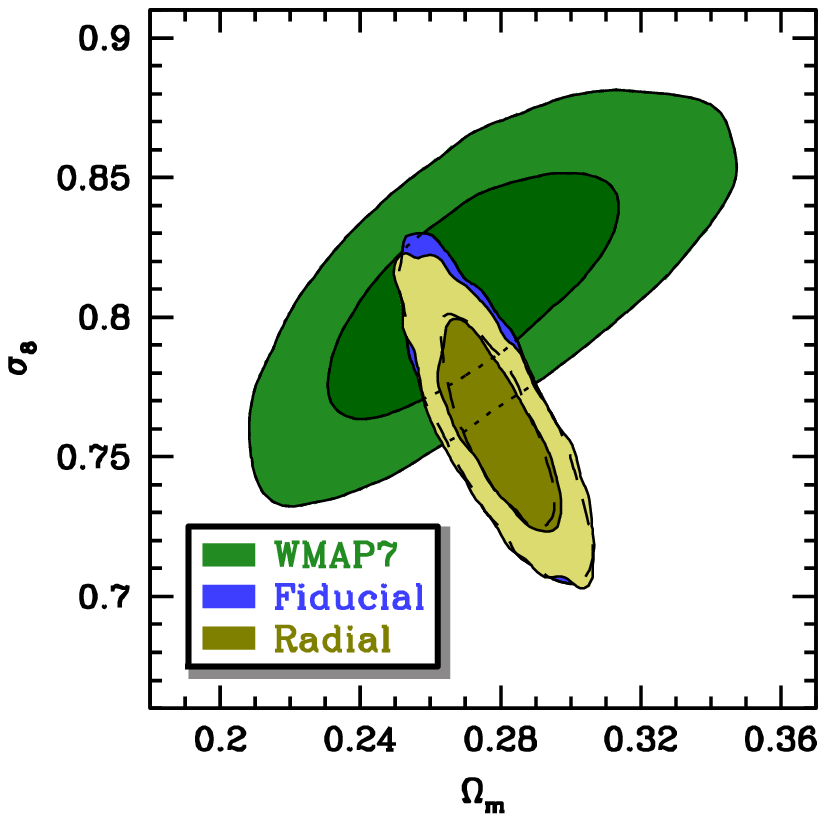,width=0.5\hdsize} &
\psfig{figure=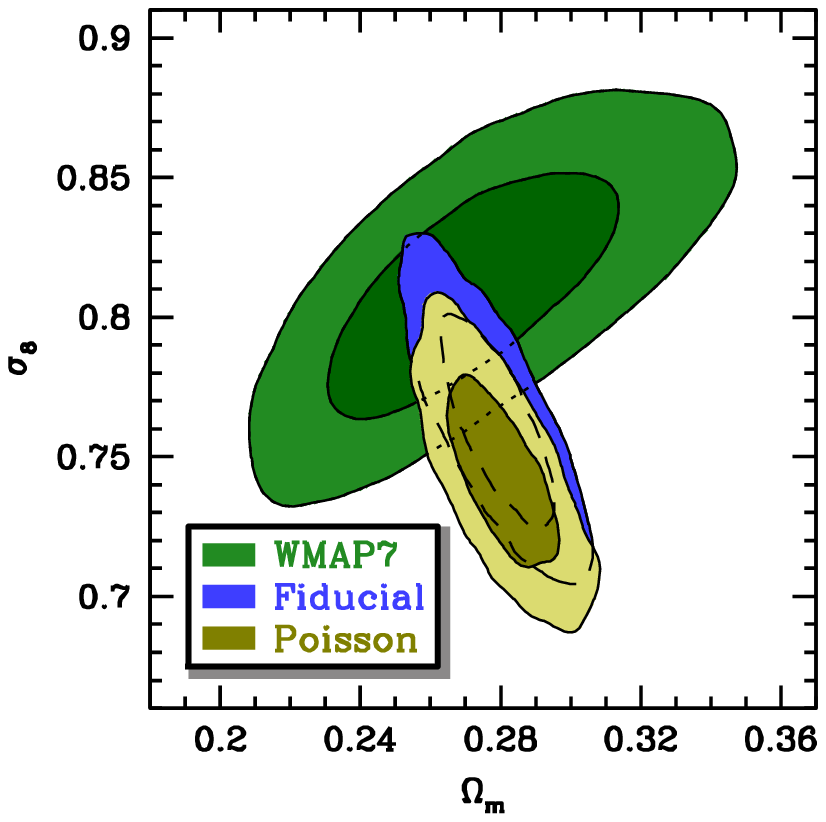,width=0.5\hdsize} \\
\psfig{figure=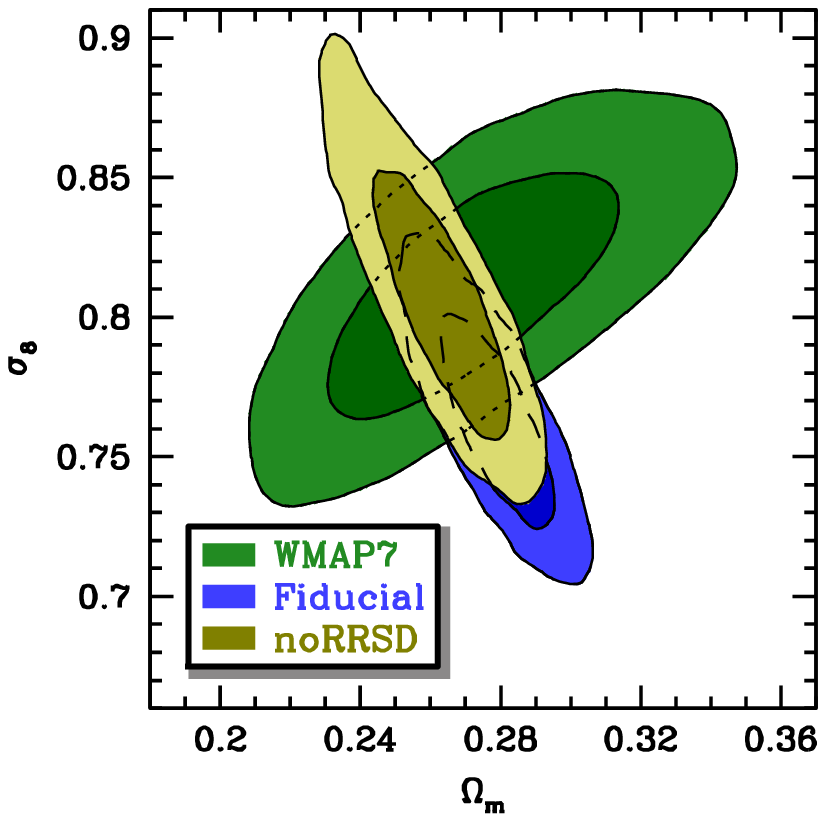,width=0.5\hdsize} &
\psfig{figure=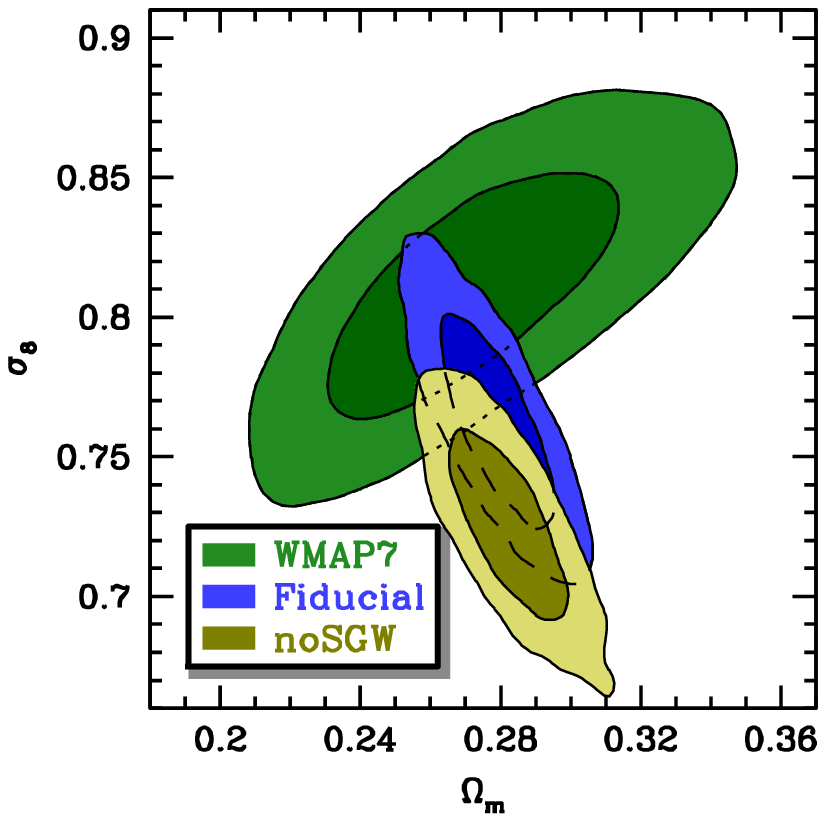,width=0.5\hdsize} 
\end{array}$
\end{center}
\caption{A comparison of the 68\% and 95\% CLs of the joint,
  two-dimensional marginalized posterior distributions of 
  $\Omega_\rmm$ and $\sigma_8$
  for our fiducial model in blue contours and, in yellow contours,
  model \modG0 (upper left-hand panel), model \modD10 (upper
  right-hand panel), model \modJ0b (lower left-hand panel), and model
  \modK0 (lower right-hand panel). In all panels the green contours
  correspond to the constraints from WMAP7, and are shown for
  comparison. See text for detailed discussion and Table~4 for a
  listing of the corresponding constraints on the cosmological
  parameters.}
\label{fig:alto}
\end{figure*}

The upper right-hand panel of Fig.~\ref{fig:alto} shows the 68 and
95\% confidence levels of the joint, two-dimensional marginalized posterior
distribution of $\Omega_\rmm$ and $\sigma_8$ for model \modD10 (yellow
contours), compared to that for our \modF0 model (blue contours). As
is evident, the extra freedom in $\calA_\rmp$ results in a best-fit
value for $\sigma_8$ that is slightly lower than for the \modF0 model
(see Table~4), but the change is only marginally significant (i.e.,
$<1\,\sigma$).

\begin{figure}
\centerline{\psfig{figure=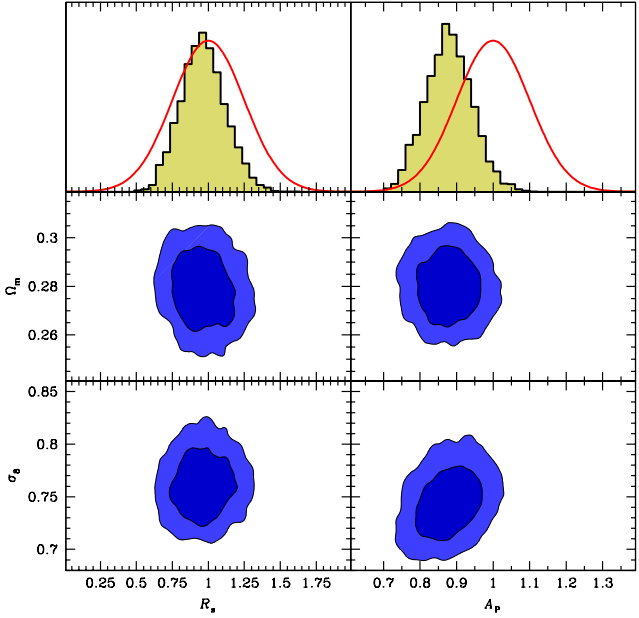,width=\hssize}}
\caption{The shaded histograms in the upper panels show the
  marginalized posterior distributions for the parameter $\calR$
  (left-hand panel), which controls the radial distribution of
  satellite galaxies (see Eq.~[\ref{defcalR}]), and $\calA_\rmP$
  (right-hand panel), which describes deviations of the satellite
  occupation distribution from a Poisson distribution (see
  Eq.~[\ref{defAp}]).  The solid, red curves show the corresponding
  Gaussian prior distributions used in models \modG0 and \modD10,
  respectively.  The contour plots in the middle and lower panels show
  the joint, two-dimensional marginalized posterior distributions of
  $\Omega_\rmm$ and $\calR$ (middle left-hand panel), $\sigma_8$ and
  $\calR$ (lower left-hand panel), $\Omega_\rmm$ and $\calA_\rmP$
  (middle right-hand panel) and $\sigma_8$ and $\calA_\rmP$ (lower
  right-hand panel). See \S\ref{sec:nrsat} and \S\ref{sec:nonpoisson}
  for details.}
\label{fig:CosmoGal}
\end{figure}

The marginalized posterior distribution for $\calA_\rmP$, shown as the
shaded histogram in the upper right-hand panel of
Fig.~\ref{fig:CosmoGal}, has $\calA_\rmP = 0.88_{-0.12}^{+0.13}$ (95\%
CL), indicating that the data prefers a sub-Poisson probability
distribution $P(N_\rms|M)$. The middle and bottom right-hand panels of
Fig.~\ref{fig:CosmoGal} show the 68 and 95 \% confidence levels of the
joint, two-dimensional posterior distributions for $\calA_\rmP$ and
$\Omega_\rmm$ (middle row) and for $\calA_\rmP$ and $\sigma_8$ (bottom
row). Whereas $\Omega_\rmm$ is found to be completely independent of
$\calA_\rmp$, there is a weak, positive correlation between $\calA_\rmP$
and $\sigma_8$ (cross-correlation coefficients $r \simeq 0.3$). Hence,
if we would have adopted a less restrictive prior on $\calA_\rmP$ the
data most likely would have preferred an even smaller value for
$\sigma_8$. We emphasize, though, that our prior has both empirical
and theoretical motivation. Nevertheless, it is clear that tighter
constraints on $\sigma_8$ require a more accurate assessment of the
(non)-Poisson nature of $P(N_\rms|M)$.

\subsection{Residual Redshift Space Distortions}
\label{sec:noRSD}

In \S\ref{sec:wpesd} we have shown that, when computing the projected
galaxy correlation functions, we use a modified form of the `Kaiser'
formalism (Kaiser 1987) to account for the residual redshift space
distortions (hereafter RRSD) present in the data. These RRSD arise as
a consequence of the fact that the line-of-sight projection was only
performed out to a distance of $r_{\rm max} = 40 h^{-1}\Mpc$ or $60
h^{-1} \Mpc$ (depending on the magnitude bin, see Table~2).  In
Paper~I, we have shown that not taking this correction into account
results in systematics errors that can easily exceed 20 percent on
scales above $\sim 10$ Mpc. In this section, we show how not
correcting for RRSD can bias the inference of cosmological parameters
such as $\Omega_{\rm m}$ and $\sigma_8$. To this aim, we run a MCMC in
which we compute the projected galaxy correlation function without
correcting for RRSD, i.e., we use
\begin{equation}\label{wpnorssd}
w_\rmp(r_\rmp) = 2 \int_{r_\rmp}^{r_{\rm out}} \xi_{\rm gg}(r) \, 
{r \, \rmd r \over \sqrt{r^2 - r_\rmp^2}}\,,
\end{equation}
with $r_{\rm out} = \sqrt{r_\rmp^2 + r^2_{\rm max}}$. As discussed in
Paper~I, this is the standard method used by numerous authors in the
past (e.g., Zehavi \etal 2004, 2005, 2011; Abazajian \etal 2005;
Tinker \etal 2005; Zheng \etal 2007, 2009; Yoo \etal 2009).  In what
follows we refer to this MCMC as our \modJ0b model.

The lower left-hand panel of Fig.~\ref{fig:alto} shows the 68\% and
95\% confidence levels of the joint, two-dimensional marginalized posterior
distribution of $\Omega_\rmm$ and $\sigma_8$ for model \modJ0b (yellow
contours), compared to that for our \modF0 model (blue contours) As is
evident, not correcting for the RRSD biases $\sigma_8$ high by $\sim
6$\% and $\Omega_\rmm$ low by about $\sim 5$\% (cf. Table~4).
Although this systematic bias is only at the $\sim 1\,\sigma$ level
for the current data, its significance will increase when using more
accurate data and/or data that extends to larger radii. Hence, we
conclude that, in this era of precision cosmology, properly correcting
for RRSD is important.
\begin{table*}\label{tab:CLFres}
\caption{Constraints on CLF Parameters}
\begin{tabular}{cccccccccc}
\hline\hline
Model & $\log(M_1)$ & $\log(L_0)$ & $\gamma_1$ & $\gamma_2$ & $\sigma_\rmc$ & $\alpha_\rms$ & $b_0$ & $b_1$ & $b_2$ \\
  (1) & (2) & (3) & (4) & (5) & (6) \\
\hline
\modF0  & $11.24_{-0.12}^{+0.12}$ & $9.95_{-0.05}^{+0.05}$ & $3.18_{-0.57}^{+0.80}$ & $0.245_{-0.012}^{+0.011}$ & $0.157_{-0.007}^{+0.007}$ & $-1.18_{-0.05}^{+0.05}$ & $-1.17_{-0.42}^{+0.28}$ & $1.53_{-0.32}^{+0.46}$ & $-0.217_{-0.122}^{+0.091}$ \\
\modG0  & $11.24_{-0.13}^{+0.11}$ & $9.95_{-0.05}^{+0.04}$ & $3.17_{-0.52}^{+0.87}$ & $0.245_{-0.011}^{+0.011}$ & $0.157_{-0.007}^{+0.007}$ & $-1.17_{-0.05}^{+0.05}$ & $-1.c14_{-0.55}^{+0.30}$ & $1.50_{-0.31}^{+0.54}$ & $-0.212_{-0.133}^{+0.086}$ \\
\modD10 & $11.23_{-0.12}^{+0.11}$ & $9.95_{-0.05}^{+0.04}$ & $3.30_{-0.53}^{+0.83}$ & $0.247_{-0.011}^{+0.012}$ & $0.158_{-0.006}^{+0.007}$ & $-1.18_{-0.05}^{+0.05}$ & $-1.15_{-0.38}^{+0.28}$ & $1.52_{-0.33}^{+0.43}$ & $-0.210_{-0.117}^{+0.095}$ \\
\modJ0b & $11.23_{-0.13}^{+0.11}$ & $9.96_{-0.05}^{+0.04}$ & $3.00_{-0.45}^{+0.78}$ & $0.239_{-0.012}^{+0.012}$ & $0.158_{-0.007}^{+0.007}$ & $-1.17_{-0.05}^{+0.05}$ & $-1.28_{-0.46}^{+0.32}$ & $1.68_{-0.36}^{+0.51}$ & $-0.262_{-0.132}^{+0.099}$ \\
\modK0  & $11.02_{-0.10}^{+0.16}$ & $9.86_{-0.04}^{+0.06}$ & $4.77_{-1.18}^{+0.75}$ & $0.273_{-0.015}^{+0.014}$ & $0.142_{-0.009}^{+0.010}$ & $-1.01_{-0.12}^{+0.11}$ & $-0.64_{-0.45}^{+0.23}$ & $1.17_{-0.27}^{+0.41}$ & $-0.208_{-0.108}^{+0.093}$ \\
\hline\hline
\end{tabular}
\medskip
\begin{minipage}{\hdsize}
  Constraints on the CLF parameters for the various models discussed
  in the text (see also Table~3).  Values indicate the median and 95\%
  CLs of the marginalized posterior distributions obtained from our
  MCMCs.
\end{minipage}
\end{table*}

\subsection{Sample Variance}
\label{sec:variance}

Even with a survey the size of SDSS, it is important to address the
issue of sample variance.  This requires full covariance matrices for
the entire data set (including the covariance between the LF, WP and
ESD data), and requires large numbers of realistic mock catalogs. This
in turn requires a large suite of very large and high resolution
$N$-body simulations, which we, unfortunately, do not have available
at this stage. An alternative is to use jackknife or bootstrap
estimates for the covariance matrix, which can be constructed from the
data itself. Although these fail to account for sample variance on the
scale of the survey itself, parameter uncertainties estimated from
jackknife errors are likely to be conservative (Norberg \etal
2011). Unfortunately, since we do not have access to the galaxy-galaxy
lensing data products, we cannot construct such a jackknife covariance
matrix. We have recently started a project to perform our own analysis
of the galaxy-galaxy lensing signal in the SDSS, following a method
similar to that of Mandelbaum \etal (2005, 2006). Once completed, we
will use those data together with our own measurements of the
clustering and luminosity (or stellar mass) functions (see Yang \etal
2012), to construct a full covariance matrix for our data set using
jackknife or bootstrap estimates.

In this study, however, we have ignored sample variance in our
analysis, and we caution that our error estimates for the model
parameters are therefore likely to be underestimated.  However, we can
still perform a simple test to gauge the potential impact of sample
variance on our results. As already alluded to in \S\ref{sec:results},
the most important manifestation of sample variance in the SDSS
appears to be the Sloan Great Wall (SGW). In order to get a rough
estimate of how the presence or absence of the SGW in our data sample
impacts on the cosmological constraints, we perform the following
exercise. We construct a MCMC using the same ingredients and priors as
for our fiducial model, but we exclude the $w_\rmp(r_\rmp)$ data for
the $[-20,-21]$ magnitude bin. As shown by Zehavi \etal (2011), this
is the magnitude bin that is most severely affected by the Sloan Great
Wall (SGW). We refer to the resulting MCMC as model \modK0. The lower
right-hand panel of Fig.~\ref{fig:alto} shows that excluding this part
of the data lowers the value of $\sigma_8$ by $\sim 5$\% to $\sigma_8
= 0.727_{-0.049}^{+0.054}$ (see Table~4), whereas $\Omega_\rmm$
remains largely unaffected.  Note that we only removed that subset of
the $w_\rmp(r_\rmp)$ data that is most severely affected by the SGW,
but that we did not remove the corresponding data from the ESD or LF.
The cosmological constraints obtained for model \modK0 therefore have
to be interpreted with care. The main purpose of this exercise is to
demonstrate that sample variance, even in a survey the size of SDSS,
is still likely to be the dominant source of error in the analysis
presented here.

%

\section{Conclusions}
\label{sec:conclusions}

Galaxies can be observed over cosmological scales, making them useful
tracers of the underlying dark matter density field.  Hence, their
spatial distribution is a potentially powerful probe for cosmology.
Unfortunately, galaxies are biased tracers of the mass distribution,
and this so-called galaxy bias depends on galaxy properties, on scale,
and is even know to be stochastic.  Because of these complications,
galaxy bias is generally considered a nuisance when using galaxies to
constrain cosmology. However, since galaxy bias also contains a wealth
of information regarding galaxy formation, it would be highly
advantageous if one could {\it simultaneously} solve for cosmology and
galaxy bias.

Several studies have used galaxy power spectra to simultaneously
constrain cosmology and galaxy bias (e.g., Cole \etal 2005; Tegmark
\etal 2006; Padmanabhan \etal 2007). These studies, however, often
resulted in tension regarding the cosmological matter density,
$\Omega_\rmm$. This is likely to be a consequence of the fact that
these studies used a poorly motivated, inaccurate fitting function for
the scale-dependence of the galaxy bias due to non-linearity,
causing biased estimates of the cosmological parameters (e.g.,
S\'anchez \& Cole 2008; Yoo \etal 2009; Hamann \etal 2008).  This
indicates that it is prudent to use a realistic, physically motivated
model for the galaxy bias.

In this paper we have simultaneously constrained cosmology and galaxy
bias using measurements of galaxy abundances, galaxy clustering and
galaxy-galaxy lensing, all taken from the SDSS.  We used a realistic,
accurate and intuitive model for galaxy bias, based on the conditional
luminosity function (which describes the halo occupation statistics as
function of galaxy luminosity) combined with the halo model (which
describes the non-linear matter field in terms of its halo building
blocks).  We improved on previous, similar studies (e.g., Seljak \etal
2005; Cacciato \etal 2009) by (i) using a more accurate analytical
model, (ii) using the latest clustering data, (iii) correcting the
latter for residual redshift space distortions, and (iv) marginalizing
over uncertainties related to the detailed structure of dark matter
haloes. 

As detailed in Paper~I (van den Bosch \etal 2012), 
our model, which includes scale-dependence of the
halo bias and a proper treatment of halo exclusion, reproduces the
3-dimensional galaxy-galaxy correlation and the galaxy-matter
cross-correlation (which can be projected to predict the observables)
with an accuracy better than 10 (in most cases 5) percent. 
Since this uncertainty is not systematic, the accuracy is expected to be 
much better for the projected galaxy-galaxy correlation function and the galaxy-galaxy lensing signal, 
which are the observed quantities. 
Furthermore, although we believe the halo mass function and halo bias function,
$n(M, z)$ and $b_{\rm h}(M, z)$, obtained by Tinker et al. (2008, 2010) 
to be the most accurate to date, it is important to realize that they still can carry
uncertainties that can potentially impact the results on cosmological parameters. 
It is unclear if such uncertainties affect just the mass function
normalization and not its shape. We intend to carry out a proper
investigation of this issue in future work. Throughout this paper,
we restrict ourselves to the $n(M, z)$ and $b_{\rm h}(M, z)$ 
specified above.

In order to constrain our model parameters, we have adopted a full
Bayesian framework in which we use Monte Carlo Markov Chains (MCMC) to
probe the posterior distributions. Under the assumption of a spatially
flat, vanilla $\Lambda$CDM cosmology, and using WMAP7 priors on the
secondary cosmological parameters, $(n_\rms,h,\Ombh)$, we obtain that
$\Omega_\rmm = 0.278_{-0.026}^{+0.023}$ and $\sigma_8 =
0.763_{-0.049}^{+0.064}$ (both 95\% CL). These are among the tightest
constraints on these two cosmological parameters ever obtained, and
are perfectly consistent with the constraints from WMAP7 (Komatsu
\etal 2011). We emphasize that no priors were used for either
$\Omega_\rmm$ or $\sigma_8$.  Our analysis, which uses a realistic
model for galaxy bias, therefore demonstrates that the observed
distribution of galaxies, down to the smallest non-linear scales, is
perfectly consistent with the vanilla $\Lambda$CDM cosmology, and that
the use of a realistic and accurate model for galaxy bias removes the
tension between the spatial distribution of galaxies and the CMB
anisotropies that hampered previous studies based on galaxy power
spectra.

In order to assess the robustness of our results, we have performed a
number of tests in which we relaxed some of our model assumptions.  In
particular, we have shown that the cosmological constraints are
extremely robust to assumptions regarding the radial number density
profile of satellite galaxies, and that the data is perfectly
consistent with satellite galaxies being an unbiased tracer of the
mass distribution within their host haloes. Our results are also
robust to assumptions regarding the second moment of the satellite
occupation distribution, $P(N_\rms|M)$. In our fiducial model we have
made the customary assumption that $P(N_\rms|M)$ follows a Poisson
distribution. Relaxing this assumption lowers the best-fit value for
$\sigma_8$ by $\sim 2.5$\% ($\sim 1\,\sigma$), to $\sigma_8 =
0.744_{-0.047}^{+0.056}$, while the constraints on $\Omega_\rmm$
remain almost identical. Interestingly, the data seems to have a
slight (marginally significant) preference for a $P(N_\rms|M)$ that is
sub-Poisson (i.e., for which $\langle N_\rms (N_\rms-1)|M\rangle <
\langle N_\rms|M \rangle^2$), consistent with expectations based on
the occupation statistics of (massive) dark matter subhaloes (e.g.,
Kravtsov \etal 2004; Boylan-Kolchin \etal 2010).

We have also shown that not correcting the projected correlation
functions of Zehavi \etal (2011) for residual redshift space
distortions biases $\sigma_8$ high and $\Omega_\rmm$ low, both by
$\sim 5$\%. Hence, in this era of precision cosmology, correcting for
these RRSD cannot be neglected. Fortunately, as demonstrated in
Paper~I, RRSD can be corrected for in a straightforward and accurate
manner using a slightly modified version of the Kaiser (1987)
formalism.

The Bayesian analysis presented here has not accounted for sample
variance. We have performed a simple, naive test to gauge the
potential impact of sample variance, by excluding part of the
clustering data that samples the SGW, the largest coherent structure
detected in the SDSS.  This lowers the value of $\sigma_8$ by $\sim
5$\% ($\sim 1\sigma$) compared to the fiducial model, to $\sigma_8 =
0.727_{-0.049}^{+0.054}$, whereas $\Omega_\rmm$ remains largely
unaffected. Hence, we conclude that sample variance may well be the
dominant source of uncertainty for our results. We have recently
started to develop our own pipeline to measure the galaxy-galaxy
lensing signal in the SDSS, which we will use to estimate the full
covariance for our data set using jackknife or bootstrap sampling.
This will allow us to better address the issue of sample variance for
the results presented here.


\section{Acknowledgments}

The work presented in this paper has greatly benefited from
discussions with Alexie Leauthaud, Rachel Mandelbaum, Jeremy Tinker,
Risa Wechsler, Idit Zehavi and Zheng Zheng.  MC acknowledges support
from the German-Israeli Foundation (GIF) I-895-207.7/2005, and from the 
Minerva Fellowship (by Max Planck Gesellschaft).  FvdB
acknowledges support from the Lady Davis Foundation for a Visiting
Professorship at Hebrew University. This research was supported in
part by the National Science Foundation under Grant No. NSF
PHY11-25915 and NSF PHY-0551142.



\appendix

\section{The Correlation Function of Dark Matter Haloes}
\label{app:Qk}

The function $Q(k|M_1,M_2,z)$, which enters in the 2-halo terms of the
galaxy-galaxy and galaxy-matter power spectra (Eq.~[\ref{P2h}]), is
related to the power spectrum of haloes of masses $M_1$ and $M_2$ at
redshift $z$. Taking account of the large scale bias of haloes, its
radial dependence (in the quasi-linear regime), and the fact that dark
matter haloes have a finite extent and therefore have a minimum
separation $r_{\rm min} = r_{\rm min}(M_1,M_2)$ (i.e., the so called
halo exclusion), the two-point correlation function between haloes of
masses $M_1$ and $M_2$ can be written as
\begin{eqnarray}\label{xihh}
\lefteqn{1+\xi_{\rm hh}(r,z|M_1,M_2) =} \\
& & \left[ 1 + b_\rmh(M_1,z) \, b_\rmh(M_2,z) \, \zeta(r,z) \, 
\xi_{\rm mm}(r,z) \right] \, \Theta(r-r_{\rm min})\,. \nonumber 
\end{eqnarray}
Here $b_\rmh(M,z)$ is the linear halo bias, $\zeta(r,z)$ describes the
scale dependence of the halo bias in the quasi-linear regime, and
$\Theta(x)$ is the Heaviside step function, which assures that
$\xi_{\rm hh}(r,z|M_1,M_2) = -1$ for $r < r_{\rm min}$ in order to
account for halo exclusion.  Throughout we set $r_{\rm min}={\rm
  max}[r_{200}(M_1),r_{200}(M_2)]$, in accordance with the particular
halo exclusion adopted by Tinker \etal (2008,2010), whose halo mass
function and halo bias function we use. As detailed in Paper~I, this
implies that
\begin{eqnarray}\label{QkM}
\lefteqn{Q(k|M_1,M_2,z) \equiv } \nonumber \\
& & 4 \pi \int_{r_{\rm min}}^{\infty}  \left[ 1+
\xi_{\rm hh}(r,z|M_1,M_2)\right] \,{\sin kr \over kr}\, r^2 \,\rmd r\,,
\end{eqnarray}

As shown by Tinker \etal (2005), the radial bias function can be
written in terms of the non-linear matter correlation function
according to
\begin{equation}\label{zetafit}
\zeta_0(r,z) = {[1 + 1.17 \, \xi_{\rm mm}(r,z)]^{1.49} 
\over [1 + 0.69 \, \xi_{\rm mm}(r,z)]^{2.09} }\,.
\end{equation}
Here the subscript 0 indicates that this empirical fitting function
was calibrated using $N$-body simulations in which the haloes were
identified using the friends-of-friends algorithm. We use a slightly
modified version of this calibration tuned to reproduce the halo-halo
and halo-matter correlation functions for halos identified by the
spherical overdensity algorithm (see Paper~I).  This modification only
affects scales smaller than a characteristic scale, $r_\psi$, such
that
\begin{equation}\label{zetamod}
\zeta(r,z)=\left\{ \begin{array}{ll}
  \zeta_0(r,z)     & \mbox{if $r\ge r_\psi$} \\
  \zeta_0(r_\psi,z) & \mbox{if $r\le r_\psi$} 
\end{array}\right. \,.
\end{equation}
The characteristic radius, $r_\psi$, is defined by
\begin{equation}\label{rpsidef}
\log\left[ \zeta_0(r_{\psi},z) \, \xi_{\rm mm}(r_{\psi},z) \right] = \psi\,.
\end{equation}
Using high resolution numerical simulations for a $\Lambda$CDM
cosmology, we found in Paper~I that $\psi=0.9$ can accurately
describe the halo-halo correlation function on small scales. Since we
cannot rule out that the calibration parameter $\psi$ depends
on cosmology, we include an uncertainty of $\sim 15$ percent on $\psi$
in our cosmological analysis (i.e., we treat $\psi$ as a free
`nuisance' parameter, using a Gaussian prior centered on $\psi = 0.9$
and with $\sigma_{\psi} = 0.15$).  Note that if Eq.~(\ref{rpsidef})
has no solution (i.e., when $\psi$ is sufficiently large), we set
$r_{\psi}=0$, which corresponds to simply using the fitting
function~(\ref{zetafit}) without modification.

\label{lastpage}

\begin{thebibliography}{}

\bibitem[]{Aba05}
Abazajian K., et al., 2005, ApJ, 625, 613

\bibitem[]{Aba09}
Abazajian K., et al., 2009, ApJS, 182, 543

\bibitem[]{DR4} 
Adelman-McCarthy J.K., et al., 2006, \apjs, 162, 38

\bibitem[]{Ast06}
Astier P., et al., 2006, A\&A, 447, 31

\bibitem[]{Bal00}
Balbi A., et al., 2000, ApJ, 545, 1

\bibitem[]{Bal10}
Baldauf T., Smith R.E., Seljak U., Mandelbaum R., 2010, Phys. Rev. D., 
81, 063531

\bibitem[]{Bee}
Beers T.C., Tonry J.L., 1986, ApJ, 300, 557

\bibitem[]{Ben07}
Benjamin J., et al., 2007, MNRAS, 381, 702

\bibitem[]{Ben11}
Benson B.A., et al., 2011, preprint (arXiv:1112.5435)

\bibitem[]{Ber02}
Berlind A.A., Weinberg D.H., 2002, ApJ, 575, 587

\bibitem[]{Biv03}
Biviano A., Girardi M., 2003, ApJ, 585, 205

\bibitem[]{Bla03a}
Blanton M.R., et al. 2003a, ApJ, 592, 819 

\bibitem[]{Bla03b}
Blanton M.R., et al. 2003b, AJ, 125, 2348

\bibitem[]{BK10}
Boylan-Kolchin M., Springel V., White S.D.M., Jenkins A., 2010,
MNRAS, 406, 896

\bibitem[]{Bullock01}
Bullock J.S., Kolatt T.S., Sigad Y., Somerville R.S., Kravtsov A.V., 
Klypin A.A., Primack J.R., Dekel A., 2001, MNRAS, 321, 559

\bibitem[]{Bur01}
Burles S., Nollett K.M., Turner M.S., 2001, ApJ, 552, 1

\bibitem[]{Bus11}
Busha M.T., Wechsler R.H., Behroozi P.S., Gerke B.F., Klypin A.A.,
Primack J.R., 2011, ApJ, 743, 11

\bibitem[]{Cac09}
Cacciato M., van den Bosch F.C., More S., Li R., Mo H.J., Yang X.,
2009, MNRAS, 394, 929
	
\bibitem[]{Cac12a}
Cacciato M., Lahav O., van den Bosch F.C., Hoekstra H., Dekel A., 2012,
preprint (arXiv:1203.2616)

\bibitem[]{Carl97}
Carlberg R.G., Yee H.K.C., Ellingson E., 1997, ApJ, 478, 462

\bibitem[]{Chen08}
Chen J., 2008, A\&A, 384, 347

\bibitem[]{Col05}
Cole S., et al., 2005, MNRAS, 362, 505

\bibitem[]{Col03}
Colless M., et al., 2003, preprint (arXiv:astro-ph/0306581)

\bibitem[]{Coo06}
Cooray A., 2006, MNRAS, 365, 842

\bibitem[]{Coo02}
Cooray A., Sheth R.K., 2002, Phys. Rep., 372, 1

\bibitem[]{DL99}
Dekel A., Lahav O., 1999, ApJ, 520, 24

\bibitem[]{Des05}
Desjacques V., Nusser A., 2005, MNRAS, 361, 1257

\bibitem[]{Die04}
Diemand J.,Moore B., Stadel J., 2004, MNRAS, 352, 535

\bibitem[]{WMAP5b}
Dunkley J., et al., 2009, ApJS, 180, 306

\bibitem[]{Eisen98}
Eisenstein D.J., Hu W., 1998, ApJ, 496, 605

\bibitem[]{ENS01}
Eke V.R., Navarro J.F., Steinmetz M., 2001, ApJ, 554, 114

\bibitem[]{Eke04}
Eke V.R., et al. (The 2dFGRS team), 2004, MNRAS, 355, 769

\bibitem[]{Fre01}
Freedman W.L., et al., 2001, ApJ, 553, 47

\bibitem[]{Fu08}
Fu L., et al., 2008, A\&A, 479, 9

\bibitem[]{Gao04}
Gao L., White S.D.M., Jenkins A., Stoehr F., Springel V., 2004, MNRAS, 355, 819

\bibitem[]{Gia08}
Giannantonio T., Scranton R., Crittenden R.G., Nichol R.C., Boughn S.P., 
Myers A.D., Richards G.T., 2008, Phys. Rev. D., 77, 123520

\bibitem[]{Gio10}
Giocoli C., Tormen G., Sheth R.K., van den Bosch F.C., 2010, MNRAS, 404, 502

\bibitem[]{Got05}
Gott J.R., Juric M., Schlegel D.J., Hoyle F., Vogeley M.S., Tegmark M., 
Bahcall N.A., Brinkmann J., 2005, ApJ, 624, 463

\bibitem[]{Guz02}
Guzik J., Seljak U., 2002, MNRAS, 325, 311

\bibitem[]{Guo12}
Guo Q., Cole S., Eke V., Frenk C., 2012, preprint (arXiv:1201.1296)

\bibitem[]{Guz00}
Guzzo L., et al., 2000, A\&A, 355, 1

\bibitem[]{Ham08}
Hamann J., Hannestad S., Melchiorri A., Wong Y.Y., 2008, JCAP, 07, 017

\bibitem[]{Has70}
Hastings W.K., 1970, Biometrika, 57, 97

\bibitem[]{Ho08}
Ho S., Hirata C.M., Padmanabhan N., Seljak U., Bahcall N., 2008, 
Phys. Rev. D., 78, 043519

\bibitem[]{jing2}
Jing Y.P., Mo H.J., B\"orner G., 1998, ApJ, 503, 37

\bibitem[]{Kai87}
Kaiser N., 1987, MNRAS, 227, 1

\bibitem[]{WMAP5a}
Komatsu E., et al., 2009, ApJS, 180, 330

\bibitem[]{WMAP7}
Komatsu E., et al., 2011, ApJS, 192, 18

\bibitem[]{Koo03}
Koopmans L.V.E., Treu T., Fassnacht C.D., Blandford R.D., Surpi G., 
2003, ApJ, 599, 70

\bibitem[]{Kow08}
Kowalski M., et al., 2008, ApJ, 686, 749

\bibitem[]{Kra04}
Kravtsov A.V., Berlind A.A., Wechsler R.H., Klypin A.A., Gottl\"ober
S., Allgood B., Primack J.R., 2004, ApJ, 609, 35

\bibitem[]{Lan01}
Lange A.E., et al., 2001, Phys. Rev. D., 63, 042001

\bibitem[]{Lar11}
Lares M., Lambas D.G., Dom\'ınguez M.J., 2011, AJ, 142, 13

\bibitem[]{Li09}
Li R., Mo H.J., Zuhui F., Cacciato M., van den Bosch F.C., Yang X., 
More S., 2009, MMRAS, 394, 1016

\bibitem[]{Lin04}
Lin Y.-T., Mohr J.J., Stanford S.A., 2004, ApJ, 610, 745

\bibitem[]{Lin11}
Lin H., et al., 2011, preprint (arXiv:1111.6622)

\bibitem[]{Lor94}
Lorrimer S.J., Frenk C.S., Smith R.M., White S.D.M., Zaritsky D., 1994, 
MNRAS, 269, 696

\bibitem[]{Ma00}
Ma C.-P., Fry J.N., 2000, ApJ, 543, 503

\bibitem[]{Mac07}
Macci\`o A.V., Dutton A.A., van den Bosch F.C., Moore B., Potter D., 
Stadel J., 2007, MNRAS, 378, 55

\bibitem[]{Man05}
Mandelbaum R., et al., 2005, MNRAS, 368, 715

\bibitem[]{Man06}
Mandelbaum R., Seljak U., Kauffmann G., Hirata C.M., Brinkmann J.,
2006, MNRAS, 368, 715

\bibitem[]{Mas06}
Masjedi M., et al., 2006, ApJ, 644, 54

\bibitem[]{Mat99}
Mather J.C., Fixsen D.J., Shafer R.A., Mosier C., Wilkinson D.T., 1999, 
ApJ, 512, 511

\bibitem[]{McD05}
McDonald P., et al., 2005, ApJ, 635, 761

\bibitem[]{Met53}
Metropolis N., Rosenbluth A.W., Rosenbluth M.N., Teller A.H., Teller E., 1953,
Chem. Phys., 21, 1087

\bibitem[]{Mo96}
Mo H.J., White S.D.M., 1996, MNRAS, 282, 347

\bibitem[]{MBW10}
Mo H.J., van den Bosch F.C., White S.D.M., 2010, Galaxy Formation and Evolution,
Cambridge University Press, Cambridge

\bibitem[]{Mor11b}
More S., 2011, ApJ, 741, 19

\bibitem[]{Mor09}
More S. van den Bosch F.C., Cacciato M., Mo H.J., Yang X., Li R., 2009,
MNRAS, 392, 801

\bibitem[]{Mor11}
More S. van den Bosch F.C., Cacciato M., Skibba R,. Mo H.J., Yang X., 2011,
MNRAS, 410, 210

\bibitem[]{Mor12a}
More S. van den Bosch F.C., Cacciato M., More A., Mo H.J., Yang X., 2012a,
preprint (arXiv:1204.0786)

\bibitem[]{Mor12b}
More S. van den Bosch F.C., Cacciato M., More A., Mo H.J., Yang X., 2012b,
submitted (Paper~II)

\bibitem[]{Mos10}
Moster B.P., Somerville R.S., Maulbetsch C., van den Bosch F.C., 
Maccio A.V., Naab T., Oser L., 2010, ApJ, 710, 903

\bibitem[]{NFW}
Navarro J.F., Frenk C.S., White S.D.M., 1997, ApJ, 490, 493

\bibitem[]{Net02}
Netterfield C.B., et al., 2002, ApJ, 571, 604

\bibitem[]{Ney52}
Neyman J., Scott E.L., 1952, ApJ, 116, 144

\bibitem[]{Nor01}
Norberg P., et al., 2001, MNRAS, 328, 64

\bibitem[]{Nor02}
Norberg P., et al., 2002, MNRAS, 332, 827

\bibitem[]{Nor09}
Norberg P., Baugh C.M., Gaztanaga E., Croton D.J., 2009, MNRAS, 396, 19

\bibitem[]{Nor11}
Norberg P., Gaztanaga E., Baugh C.M., Croton D.J., 2011, MNRAS, 418, 2435

\bibitem[]{Ogu08}
Oguri M., et al., 2008, AJ, 135, 512

\bibitem[]{OMe06}
O'Meara J.M., Burles S., Prochaska J.X., Prochter G.E., Bernstein R.A., 
Burgess K.M., 2006, ApJ, 649, L61

\bibitem[]{Pad07}
Padmanabhan N., White M., Eisenstein D.J., 2007, MNRAS, 376, 1702

\bibitem[]{Pad07}
Padmanabhan N., et al., 2007, MNRAS, 378, 852

\bibitem[]{Par94}
Park C., Vogeley M.S., Geller M.J., Huchra J.P., 1994, ApJ, 431, 569

\bibitem[]{john2}
Peacock J.A., Smith R.E., 2000, MNRAS, 318, 1144

\bibitem[]{Per07}
Percival W.J., et al., 2007, ApJ, 657, 645

\bibitem[]{Perl99}
Perlmutter S., et al., 1999, ApJ, 517, 565

\bibitem[]{por}
Porciani C., Magliocchetti M., Norberg P., 2004, MNRAS, 355, 1010

\bibitem[]{Pry02}
Pryke C., Halverson N.W., Leitch E.M., Kovac J., Carlstrom J.E., 
Holzapfel W.L., Dragovan M., 2002, ApJ, 568, 46

\bibitem[]{Rei09}
Reichardt C.L., et al., 2009, ApJ, 694, 1200 

\bibitem[]{RSB09}
Reid B.A., Spergel D.N., Bode P., 2009, ApJ, 702, 249

\bibitem[]{Rie98}
Riess A.G., et al., 1998, AJ, 116, 1009

\bibitem[]{Roz10}
Rozo E., et al., 2010, ApJ, 708, 645

\bibitem[]{Ruhl03}
Ruhl J.E., et al. 2003, ApJ, 599, 786

\bibitem[]{Sal05}
Sales L., Lambas D.G., 2005, MNRAS, 356, 1045

\bibitem[]{San08}
S\'anchez A.G,, Cole S., 2008, MNRAS, 385, 830

\bibitem[]{San06}
Sandage A., Tammann G.A., Saha A., Reindl B., Macchett F.D., 
Panagia N., 2006, ApJ, 653, 843

\bibitem[]{Sco01}
Scoccimarro R., Sheth R.K., Hui L., Jain B., 2001, ApJ, 546, 20

\bibitem[]{Sco94}
Scott D., Srednicki M., White M., 1994, ApJ, 421, L5

\bibitem[]{Sel00}
Seljak U., 2000, MNRAS, 318, 203

\bibitem[]{Sel05}
Seljak U., et al., 2005, Phys. Rev. D, 71, 043511

\bibitem[]{She09}
Sheldon E.S., et al., 2009, ApJ, 703, 2232

\bibitem[]{Ski11}
Skibba R.~A., van den Bosch F.~C., Yang X., More S., Mo H., Fontanot F.,
2011, MNRAS, 410, 417

\bibitem[]{Smith03}
Smith R.E., et al., 2003, MNRAS, 341, 1311

\bibitem[]{WMAP1}
Spergel D.N., et al., 2003, ApJS, 148, 175

\bibitem[]{WMAP3}
Spergel D.N., et al., 2007, ApJS, 170, 377

\bibitem[]{Spr08}
Springel V., et al., 2008, MNRAS, 391, 1685

\bibitem[]{Tal12}
Tal T., Wake D.A., van Dokkum P.G., 2012, ApJ, 751, 5

\bibitem[]{Taletal12}
Tal T., Wake D.A., van Dokkum P.G., van den Bosch F.C., Schneider D.P., 
Brinkman J., Weaver B.A., 2012, ApJ, 746, 138

\bibitem[]{Teg99}
Tegmark M., Bromley B.C., 1999, ApJ, 518, 69  

\bibitem[]{Teg06}
Tegmark M., et al., 2006, Phys. Rev. D., 74, 123507

\bibitem[]{Tin05}
Tinker J.L., Weinberg D.H., Zheng Z., Zehavi I., 2005, \apj, 631, 41

\bibitem[]{TNWW07}
Tinker J.L., Norberg P., Weinberg D.H., Warren M.S., 2007, \apj, 659, 877

\bibitem[]{Tin08a}
Tinker J.L., Kravtsov A.V., Klypin A., Abazajian K., Warren M., Yepes G., 
Gottl\"ober S., Holz D.E., 2008, \apj, 688, 709

\bibitem[]{Tin10}
Tinker J.L., Robertson B.E., Kravtsov A.V., Klypin A., Warren M., Yepes G., 
Gottl\"ober S., 2010, ApJ, 724, 878

\bibitem[]{Tin12}
Tinker J.L., et al., 2012, ApJ, 745, 16

\bibitem[]{BYM03}
van den Bosch F.C., Yang X., Mo H.J., 2003a, MNRAS , 340, 771

\bibitem[]{BMY03}
van den Bosch F.C., Mo H.J., Yang X., 2003b, MNRAS , 345, 923

\bibitem[]{vdB05}
van den Bosch F.C., Tormen G., Giocoli C., 2005a, MNRAS, 359, 1029

\bibitem[]{BYMN05}
van den Bosch F.C., Yang X., Mo H.J., Norberg P., 2005b, MNRAS, 356, 1233

\bibitem[]{vdB07}
van den Bosch F.C., Yang X., Mo H.J., Weinmann S.M., Macc\`io A.V,, 
More S., Cacciato M., Skibba R., Kang, X., 2007, MNRAS, 376, 841

\bibitem[]{vdB12}
van den Bosch F.C., More S., Cacciato M., Mo H.J., Yang X., 2012, 
preprint (arXiv:1206.6890), (Paper~I)

\bibitem[]{vdM00}
van der Marel R.P., Magorrian J., Carlberg R.G., Yee H.K.C., Ellingson
E., 2000, AJ , 119, 2038

\bibitem[]{vanL07}
van Leeuwen F., Feast M.W., Whitelock P.A., Laney C.D., 2007, MNRAS, 379, 723

\bibitem[]{Vie04}
Viel M., Weller J., Haehnelt M.G., 2004, MNRAS, 355, L23

\bibitem[]{Vik09}
Vikhlinin A., et al., 2009, ApJ, 692, 1060

\bibitem[]{Wan06}
Wang L., Li C., Kauffmann G., De Lucia G., 2006, MNRAS, 371, 537

\bibitem[]{Wan07}
Wang Y., Yang X., Mo H.J., van den Bosch F.C., 2007, ApJ, 664 608

\bibitem[]{Wat10}
Watson D.F., Berlind A.A., McBride C.K., Masjedi M., 2010, ApJ, 709, 115

\bibitem[]{Wat12}
Watson D.F., Berlind A.A., McBride C.K., Hogg D.W., Jiang T., 2012, ApJ, 
749, 83

\bibitem[]{YMB03}
Yang X., Mo H.J., van den Bosch F.C., 2003, MNRAS, 339, 1057

\bibitem[]{Yang04}	
Yang X. Mo H.J., Jing Y.P., van den Bosch F.C., Chu Y., 2004, MNRAS, 350, 1153
        
\bibitem[]{Yang05}
Yang X., Mo H.J., van den Bosch F.C., Weinmann S.M., Li C., Jing Y.P.,
2005, MNRAS, 362, 711

\bibitem[]{Y07} 
Yang X., Mo H.J., van den Bosch F.C., Pasquali A., Li C., Barden M.,
2007, ApJ, 671, 153 

\bibitem[]{YMB08}
Yang X., Mo H.J., van den Bosch F.C., 2008, ApJ, 676, 248

\bibitem[]{YMB12}
Yang X., Mo H.J., van den Bosch F.C., Zhang Y., Han J., 2012, ApJ, 752, 41

\bibitem[]{Yoo06}
Yoo J., Tinker J.L., Weinberg D.H., Zheng Z., Katz N., Dav\'e R., 2006, ApJ, 
652, 26

\bibitem[]{Yoo09}
Yoo J., Weinberg D.H., Tinker J.L., Zheng Z., Warren M.S., 2009, ApJ, 698, 967

\bibitem[]{York00}
York D.G., et al., 2000, AJ, 120, 1579

\bibitem[]{Zeh04}
Zehavi I., et al., 2004, ApJ, 608, 16

\bibitem[]{Zeh05}
Zehavi I., et al., 2005, ApJ, 630, 1

\bibitem[]{Zeh11}
Zehavi I., et al., 2011, ApJ, 736, 59

\bibitem[]{Zhao09}
Zhao D.H., Jing Y.P., Mo H.J., B\"orner G., 2009, ApJ, 707, 354

\bibitem[]{Zhe07}
Zheng Z., Coil A., Zehavi I., 2007, ApJ, 667, 760

\bibitem[]{Zhe09}
Zheng Z., Zehavi I., Eisenstein D.J., Weinberg D.H., Jing Y.P., 2009, ApJ, 
707, 554

\end{thebibliography}
\end{document}